\documentclass{kais}
\usepackage{amsmath}
\usepackage{amssymb}
\usepackage{graphicx}
\usepackage{subfigure}

\received{xxx}
\revised{xxx}
\accepted{xxx}

\pubyear{2000}
\pagerange{\pageref{firstpage}--\pageref{lastpage}}
\volume{xxx}

\graphicspath{{../figs/},{../figs-new/}}

\def\Hline{\noalign{\hrule height 4\arrayrulewidth}}
\def\fline{\noalign{\hrule height 1\arrayrulewidth}}

\begin{document}

\author[A. Beykikhoshk et al]{Adham Beykikhoshk$^\dag$, Ognjen Arandjelovi\'c$^\ddag$, Dinh Phung$^\dag$, Svetha Venkatesh$^\dag$\\
$^\dag$ Pattern Recognition and Data Analytics Centre, Deakin University, Australia \\
\{abeyki, dinh.phung, svetha.venkatesh\}@deakin.edu.au\vspace{10pt}\\
$^\ddag$ School of Computer Science, University of St Andrews, UK\\
ognjen.arandjelovic@gmail.com}

\title{Discovering Topic Structures of a Temporally Evolving Document Corpus}

\maketitle

\begin{abstract}
In this paper we describe a novel framework for the discovery of the topical content of a data corpus, and the tracking of its complex structural changes across the temporal dimension. In contrast to previous work our model does not impose a prior on the rate at which documents are added to the corpus nor does it adopt the Markovian assumption which overly restricts the type of changes that the model can capture. Our key technical contribution is a framework based on (i) discretization of time into epochs, (ii) epoch-wise topic discovery using a hierarchical Dirichlet process-based model, and (iii) a temporal similarity graph which allows for the modelling of complex topic changes: emergence and disappearance, evolution, splitting, and merging. The power of the proposed framework is demonstrated on two medical literature corpora concerned with the autism spectrum disorder (ASD) and the metabolic syndrome (MetS) -- both increasingly important research subjects with significant social and healthcare consequences. In addition to the collected ASD and metabolic syndrome literature corpora which we made freely available, our contribution also includes an extensive empirical analysis of the proposed framework. We describe a detailed and careful examination of the effects that our algorithms's free parameters have on its output, and discuss the significance of the findings both in the context of the practical application of our algorithm as well as in the context of the existing body of work on temporal topic analysis. Our quantitative analysis is followed by several qualitative case studies highly relevant to the current research on ASD and MetS, on which our algorithm is shown to capture well the actual developments in these fields.


\end{abstract}

%\linespread{2}

\section{Introduction\label{s:intro}}
In the last decade and a half so-called topic modelling has emerged as a powerful statistical paradigm for the automatic semantic analysis of large collections of documents. Topic models as their name suggests can be seen as formalizations of the colloquial understanding of `topics' addressed in a piece of text. More specifically, in this context a topic becomes a probability distribution over a fixed vocabulary of words (or more generally terms). Thus an example of a particular topic may be:\\
\begin{align}
  \centering
   \underbrace{
  \begin{array}{ccccc}
      0.2         &      0.15      &     0.11              & 0.06      &   \ldots   \\
      autistic   &     child      &     education     &  needs   &  \ldots \\
   \end{array}
    }_\text{vocabulary size}
\end{align}
where the upper row contains probabilities which correspond to the vocabulary words shown in the bottom row. Using higher order semantic understanding a human interpreting this formal representation of a topic may describe it as capturing a discussion of educational needs of children with ASD although it should be noted that such interpretation may not always be straightforward~\cite{ChanGerrWangBoyd+2009}.

\begin{figure}
  \centering
  \subfigure[Example topic 1]{\fbox{\includegraphics[width=0.35\textwidth]{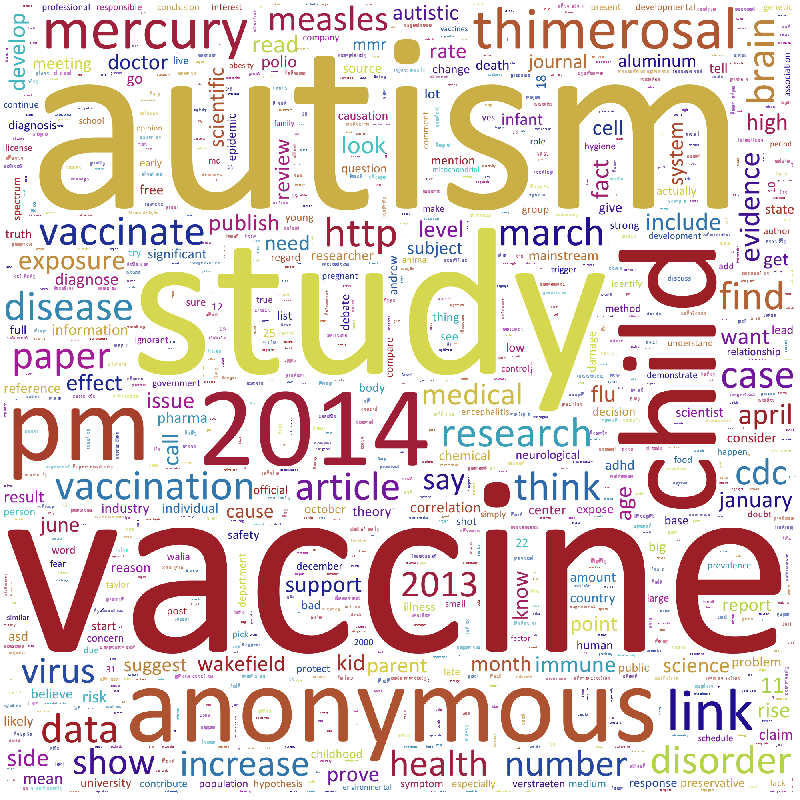}}}~~~~~
  \subfigure[Example topic 2]{\fbox{\includegraphics[width=0.35\textwidth]{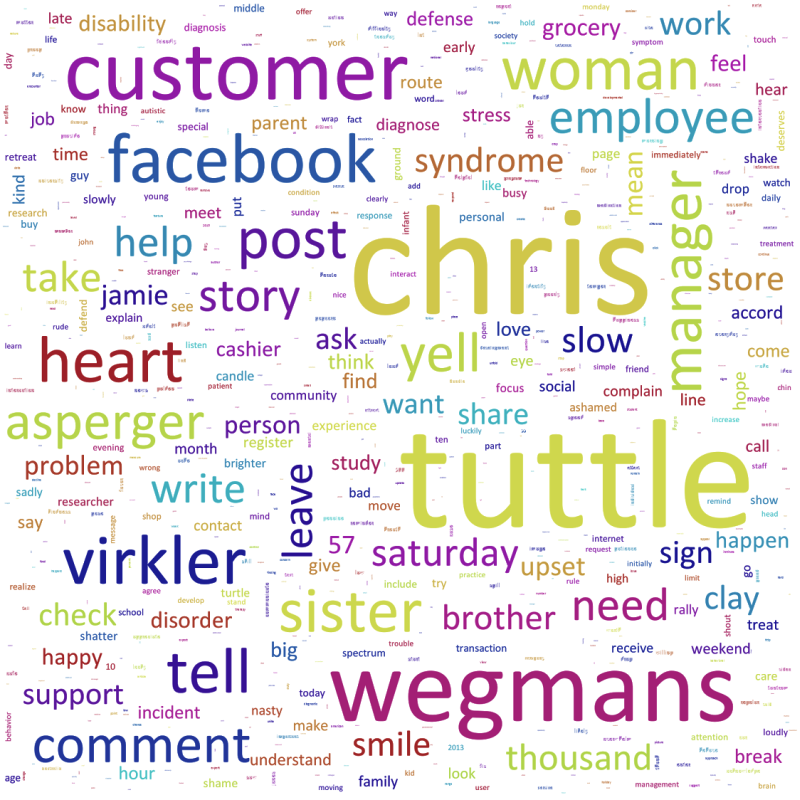}}}
  \subfigure[Example topic 3]{\fbox{\includegraphics[width=0.35\textwidth]{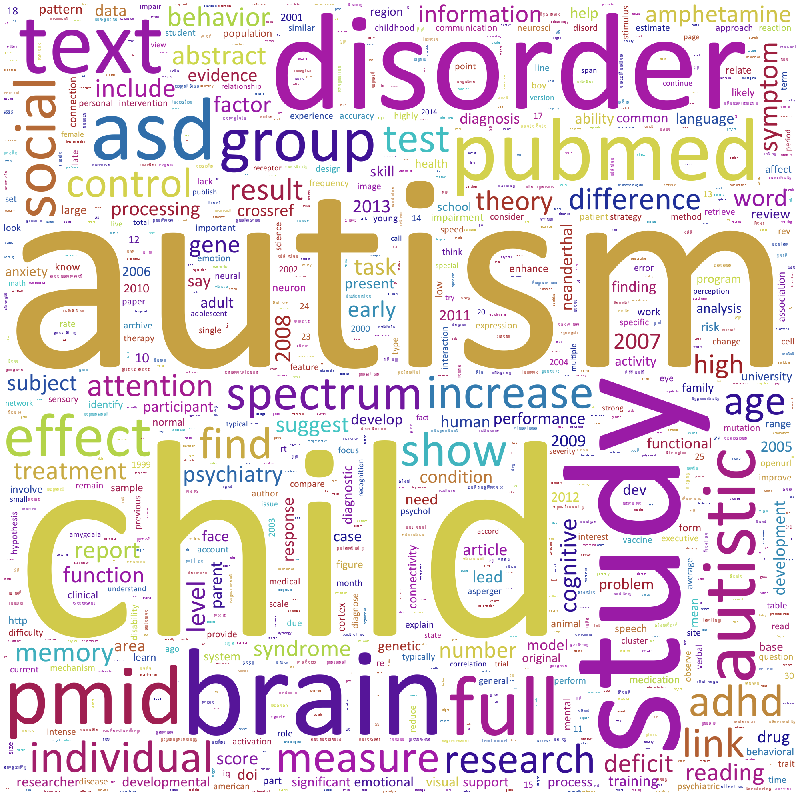}}}~~~~~
  \subfigure[Example topic 4]{\fbox{\includegraphics[width=0.35\textwidth]{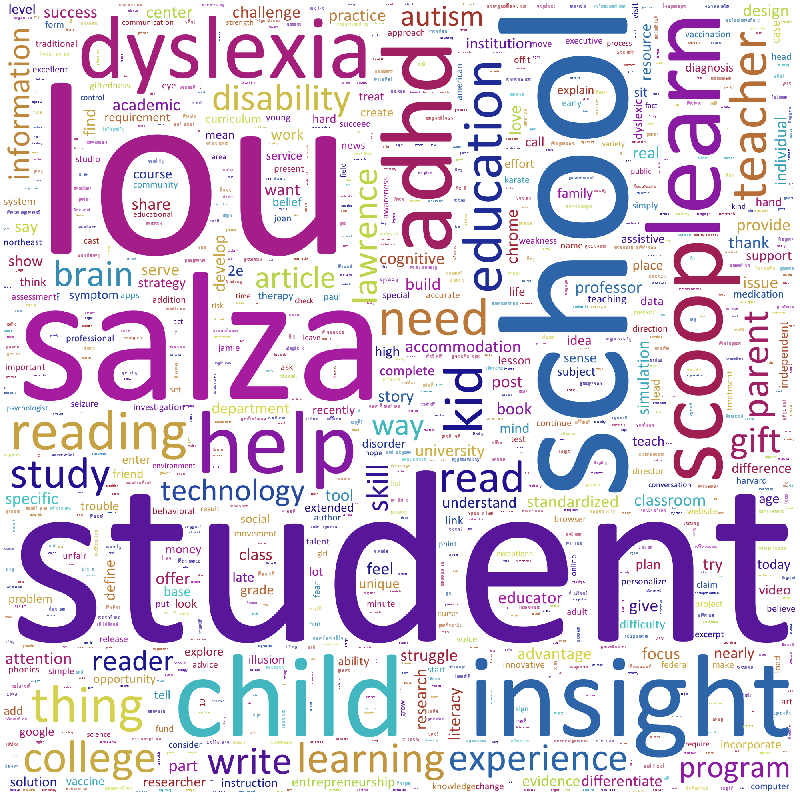}}}
  \caption{ An illustration of typical topics discovered by the proposed method in a single epoch, shown as size-coded word-clouds -- a larger font indicates a proportionally more probable term (different colours are merely used for easier visualization and encode no information pertaining to the corresponding terms themselves). Topic~1 can be readily related to the persistent myth of a link between childhood vaccination and autism development as well as related issues concerning the effects of thimerosal and mercury~\cite{HviiStelWohlMelb2003}, Topic~2 to the incident involving Chris Tuttle, a Wegmans employee
  with Asperger's syndrome who attracted media attention in November 2013, Topic~3 to medical literature on autism, children, and brain development, and Topic~4 to the schooling of children with learning disabilities. }
  \label{f:topics}
\end{figure}

Most research on topic modelling to date has focused on the analysis in the context of \emph{static} corpora, that is, document collections which do not possess a temporal structure.  In such collections documents are said to be interchangeable.  The key techniques dominating this domain are Bayesian non-parametric inference algorithms and the latent Dirichlet allocation (LDA) in particular, first described by Blei \textit{et al.}~\cite{blei2003latent} and subsequently extended in a variety of ways. Indeed at the bottom-most level the present paper uses a model based on the hierarchical Dirichlet process (HDP) which is one of the aforementioned extensions. Both LDA and HDP are explained in some detail in Section~\ref{ss:mixModels}.

However, in many problems of practical significance, it is not only the instantaneous topic structure that is of interest -- the change in this structure over time too often conveys important information and insight. For this reason in recent years the problem of temporal topic modelling has been attracting an increasing amount of research attention~\cite{BeykPhunAranVenk2015,BeykAranPhunVenk2015a}.  Indeed the focus of the present paper is on temporally changing corpora. At the heart of the method that we describe is an automatically constructed and temporally constrained graph superimposed on topics extracted from short time epochs within which the corpus can be considered as being static.

In particular the work described herein extends our contribution first described in~\cite{BeykPhunAranVenk2015} (also see \cite{BeykAranPhunVenk+2014,BeykAranPhunVenk+2015} for related prior work). Retaining the same structural framework, in the present paper we analyse in detail the effects of different pruning parameters in the construction of the graph superimposed over the extracted topics. Specifically we consider different choices of inter-topic distance measures and the process of selecting an appropriate pruning threshold given a specific distance measure. In addition, we describe extended experiments on the collection of abstracts of scholarly articles on the autism spectrum disorder (ASD), gathered by ourselves and made publicly available, as well as additional experiments on a newly gathered collection of abstracts of scholarly articles in the highly active sphere of work on the metabolic syndrome. This new collection of documents will also be made publicly available.

The remainder of the present paper is structured as follows.  In the next section we review relevant previous work, both on topic modelling in general as well as on temporal modelling which is the focus of our work.  Then in Section~\ref{s:proposed} we first describe the key prerequisite modelling techniques which our contributions builds upon (Section~\ref{ss:mixModels}), followed by the detail on the proposed modelling framework (Section~\ref{ss:contrib}). Section~\ref{ss:contrib} introduces our main technical contributions. Specifically these are our general temporal model based on what we term the temporal similarity graph, and the two key aspects involved in the construction of this graph: the choice of an inter-topic similarity measure (Section~\ref{ss:similarity}) and a graph pruning process (Section~\ref{ss:construction}). The proposed method is extensively evaluated in Section~\ref{s:eval}. Section~\ref{ss:evalData} motivates our focus on biomedical documents used in the evaluation with Section~\ref{ss:prevBiomed} summarizing previous work in the specific domain of text mining, Section~\ref{sss:asd&mets} provides a brief synopsis of the medical conditions the two specific document data sets we used for evaluation relate to, and Section~\ref{ss:data} describes the data sets themselves, the manner in which data was collected, as well as the pre-processing thereof. Section~\ref{ss:evalMeasures} details different inter-topic similarity measures we compared. Our experiments and the associated findings and discussion are presented in Section~\ref{ss:exp}, with Section~\ref{sss:quant} focusing on quantitative elements of the evaluation and Section~\ref{sss:qual} on qualitative analysis. Finally, the contributions and the key findings of the present work are summarized in Section~\ref{s:conc}.

\section{Previous work\label{s:previous}}
Owing to its growing popularity in recent years, the existing literature on probabilistic topic modelling is already vast. Hence a comprehensive survey of the field is out of the scope of the present paper. Herein we overview the broad and most influential research directions, with a particular focus on techniques of direct relevance to the methodology described in the proposed paper. In particular we direct our attention first to latent topic models which have dominated the field in the last decade, and then on biomedical text mining, given the application domain within which our framework is evaluated in Section~\ref{s:eval}.

Topic models in modern machine learning are often described as latent probabilistic models. The attribute `latent' is intended to capture the nature of inferred topics -- these are hidden variables in the sense that they are not explicit in the observable data itself. Arguably the sought after ``true'' topic structure is also neither objective nor accessible even in principle. On the other hand, the attribute `probabilistic' conveys the inherent aspect of the aforementioned models whereby modelling imperfections as well as ambiguities in data are handled through the use of probability distributions which readily accommodate uncertainty. Therefore we start our discussion with the simplest latent topic models which formalize the key ideas in the field and underlie much of the subsequent, more complex models.

\subsection{Latent topic models\label{ss:latent}}
An important early topic modelling approach comes in the form of so-called latent semantic indexing (LSI)~\cite{deerwester1990indexing} which remains popular. Two notable limitations of LSI are its inability to deal effectively with polysemy and to produce an explicit description of the latent space. A probabilistic improvement overcomes these by explicitly characterizing the latent space with semantic topics, and by employing a probabilistic generative model that addresses the polysemy problem~\cite{hofmann1999probabilistic}. Nevertheless, probabilistic LSI is prone to parameter overfitting caused by an uncontrolled growth in the number of parameters as the document corpus is increased. In addition, the necessary assignment of probabilities to documents is a nontrivial task~\cite{blei2003latent}.

The recently proposed latent Dirichlet allocation (LDA) method~\cite{blei2003latent} overcomes the overfitting problem by adopting a Bayesian framework and a generative process at the document level. While LDA has quickly become a standard tool for topic modelling, it too experiences challenges when applied on real-world data. In particular, being a parametric model the number of desired output topics has to be specified in advance. The HDP model as the nonparametric counterpart of LDA was introduced by Teh \textit{et al.}~\cite{teh_etal_2006_hierarchical} and addressed this limitation by using a hierarchical Dirichlet process (as opposed to a Dirichlet distribution) as the prior on topics. Therefore, each document is modelled using an infinite mixture model, allowing the data to inform the complexity of the model and infer the number of resulting topics automatically. We discuss this model in further detail in Section~\ref{s:proposed}.

\subsubsection{Temporal topic modelling}
A notable limitation of most models described in the existing literature lies in their assumption that the data corpus is static; this includes those based on LDA mentioned previously, or the hierarchical Dirichlet process described in detail in the next section. Here the term `static' is used to describe the lack of any associated temporal (or indeed sequential) information associated with the documents in a corpus -- the documents are said to be exchangeable~\cite{blei2006dynamic}. However, in many practical applications documents are added to the corpus in a temporal manner and their ordering has significance i.e.\ the documents are non-exchangeable.  As a consequence, the topical structure of the corpus changes over time. For example, the corpus of scholarly literature on a particular subject is a growing corpus which by its very nature exhibits significant changes in its topic structure over time~\cite{EinsInfe1961,Dyso2012}: new ideas emerge, old ideas are refined, novel discoveries result in multiple ideas being related to one another thereby forming more complex concepts or a single idea multifurcating into different `sub-ideas' which are thereafter investigated with some degree of independence etc. A good example from a different realm can be readily found in the corpus of social media contributions, such as Twitter~\cite{BeykAranPhunVenk2015a}. With an even faster pace here too complex topic changes can be observed, with novel topics of conversation being instigated by e.g.\ `real-world' events (such as epidemics, terrorist attacks, or developments in the world of popular culture), changed by the contributions of other users, split into new topics, merged with others etc.

The assumption made by all previous work, and indeed adopted by us, is that documents are not exchangeable at large temporal scales but are at short time scales, thus treating the corpus as \emph{temporally locally static}. The scale at which this assumption can be considered as valid is clearly application and corpus dependent, and is an important consideration. Indeed the present paper investigates this in detail.

The existing work on temporal topic modelling can be divided into two groups of approaches both of which can be based on parametric~\cite{blei2006dynamic,Wang2008,Wang2006_tot} or nonparametric~\cite{ren2008dynamic,zhang2010evolutionary} techniques, the former suffering from the limitation that they contain free parameters which must be set \textit{a priori}. Methods of the first group discretize time into epochs, apply a static topic model to each epoch, and by making the Markovian assumption relate the parameters of each epoch's topic model to those of the epochs adjacent to it in time~\cite{blei2006dynamic,Wang2008,ren2008dynamic,zhang2010evolutionary}. While the approach we propose in this paper adopts the idea of time discretization, it diverges in its other features from this group of methods thereafter. In particular, instead of employing the Markovian assumption we describe a novel structure in form of a temporal similarity graph, which gives our method greater flexibility, as described in detail in the next section. The second group of methods in the literature regard document time-stamps as observations of a continuous random variable~\cite{Wang2006_tot,dubey2013nonparametric}. This assumption severely limits the type of topic changes which can be described. For example, as opposed to our model, these models are not capable of describing the evolution of topics, or their splitting and merging, and are rather constrained to tracking simple topic popularity.

\section{Proposed framework\label{s:proposed}}

In this section we present the technical contribution of the present work. We begin by reviewing the relevant theory underlying Bayesian mixture modelling, LDA, and HDP that plays a central role in the proposed framework. Then we turn our attention to the novel contribution of our work and explain how the aforementioned Bayesian techniques are extended to deal with temporally varying document corpora.

% is employed to discover the topical content of a literature corpus and track its structural changes over time.

\subsection{Bayesian mixture models}\label{ss:mixModels}
In recent years mixture models have become popular choices for the modelling of so-called heterogeneous data. In this context heterogeneity is taken to mean that observable data is generated by more than one process (source). One of the key challenges in the analysis of heterogenous data lies in the lack of observability of the correspondence between specific data points and their sources i.e.\ it is not known which data source generated which data. Usually it is also the case that the number of sources is not known either~\cite{richardson1997bayesian}. Mixture models, and in particular mixture models enveloped within a Bayesian framework, have distinct advantages over alternative approaches as we shall explain shortly.

%Bayesian treatment of mixture modelling allows for including prior information in the analysis and direct probability statements about unknown variables. Moreover, it is argued that Bayesian paradigm is the only sensible approach, when the number of components is unknown~\cite{richardson1997bayesian}.

\subsubsection{Finite mixture modelling\label{sub:Finite-mixture-modelling}}
Finite mixture models rely on the assumption that the observed data is generated by $K$ clusters, each cluster being associated with the parameter $\phi_{k}$ and underlain by the probability density function $f\left(.|\phi_{k}\right)$. An observation $x$ is assumed to be generated by first choosing a cluster $k$ with probability $\pi_{k}$ followed by a random draw from the corresponding distribution described by $\phi_{k}$. Therefore the process can be summarized by the following:
\begin{align}
  p\left(x|\pi_{1:K},\phi_{1:K}\right)=\sum_{k=1}^{K}\pi_{k}f\left(x|\phi_{k}\right).
\end{align}
In a Bayesian setting the model parameters (i.e.\ mixing proportions $\pi_{1:K}$ and component parameters $\phi_{1:k}$) are further endowed by priors. Typically the symmetric Dirichlet distribution is placed on top of $\pi_{1:K}$ and a prior on $\phi_{1:K}$ conjugate with $f\left(.|\phi_{k}\right)$ chosen for computational convenience.

%Posterior inference enables us to incorporate the data and conditioned on the observed corpus, find the posterior distribution of model parameters. Thus we can estimate the density functions of the components that generated the data.

\subsubsection{Latent Dirichlet allocation\label{sss:lda}}
In the previous section we described how to model a group of data points with a Bayesian finite mixture model. Latent Dirichlet allocation adds a level of hierarchy on the mixing proportions to allow for the modelling of data points in groups that share a set of components.

Following the consensus in the literature we adopt the terminology used in the analysis of textual data (which is the context in which LDA was originally proposed~\cite{blei2003latent}) and hereafter interexchangably refer to data points as words, their groups as documents, and mixture components as topics. The technical term `topic' can be interpreted as formalizing and abstracting the colloquial notion of a topic which is understood at a higher semantic level. Therefore the modelling framework of LDA can be described by the following generative process:
\begin{align}
  \phi_{1:K} & \sim && H,\\
  \pi_j & \sim && \text{Dirichlet}\left(\alpha\right),\\
  z_{ji}|\pi_j & \sim && \pi_j,\\
  x_{ji}|z_{ji},\phi_{1:K} & \sim && F\left(\phi_{z_{ji}}\right),
\end{align}
where $H$ is the base distribution of topics, $\alpha$ the hyperparameter of the prior on the distribution of topics within a document, $\pi_j$ the distribution of topics in document $j$, and $z_{ji}$ the topic corresponding to the $i$-th word in the $j$-the document. The corresponding model likelihood is:
\begin{align}
  p\left(w_{ji}|\alpha\right)=\int_{\pi_{j}}\int_{\phi_{1:K}}\sum_{k=1}^{K}\pi_{jk}f\left(x|\phi_{k}\right)dP\left(\pi_{j}\right)dP\left(\phi_{1:K}\right),
\end{align}
Approximation techniques such as MCMC~\cite{griffiths2004finding} and Variation Bayes~\cite{blei2003latent} methods can be used for posterior inference.

\subsubsection{Infinite mixture modelling}
As mentioned earlier, LDA requires the number of topics to be fixed in advanced which is a serious limitation in practice. Choosing the number of topics is usually performed by examining how well the model fits a set of held-out documents. However, if a previously unseen topic has contributed in generating the held-out data, LDA is not able to infer correct parameters of that topic.

Bayesian non-parametric~(BNP) methods place priors on the infinite-dimensional space of probability distributions and provide an elegant solution to this problem. Dirichlet Process~(DP)~\cite{ferguson1973bayesian} as the non-parametric counterpart of
the Dirichlet distribution and the building block of BNP allows for the model to accommodate a potentially infinite number of mixture components. The generative likelihood for a data point $x$ in infinite mixture model is:
\begin{align}
p\left(x|\pi_{1:\infty},\phi_{1:\infty}\right)=\sum_{k=1}^{\infty}\pi_{k}f\left(x|\phi_{k}\right).
\end{align}

$\text{DP}\left(\gamma,H\right)$ is defined as a distribution of
a random probability measure $G$ over a measurable space $\left(\Theta,\mathcal{B}\right)$,
such that for any finite measurable partition $\left(A_{1},A_{2},\ldots,A_{r}\right)$
of $\Theta$ the random vector $\left(G\left(A_{1}\right),\ldots,G\left(A_{r}\right)\right)$
is a Dirichlet distribution with parameters $\left(\gamma H\left(A_{1}\right),\ldots,\gamma H\left(A_{r}\right)\right)$. A DP generates imperfect atomic copies of its base measure $H$ with
a variance governed by its concentration parameter $\gamma$. An alternative view of the DP emerges from the so-called stick-breaking process which adopts a constructive approach using a sequence of discrete draws~\cite{sethuraman1991constructive}. Specifically, if $G\sim\text{DP}\left(\gamma,H\right)$ then $G=\sum_{k=1}^{\infty}\beta_{k}\delta_{\phi_{k}}$
where $\phi_{k}\overset{iid}{\sim}H$ and $\boldsymbol{\beta}=\left(\beta_{k}\right)_{k=1}^{\infty}$
is the vector of weights obtained by a stick-breaking process that
is $\beta_{k}=v_{k}\prod_{l=1}^{k-1}\left(1-v_{l}\right)$ and $v_{l}\overset{iid}{\sim}\text{\text{Beta}}\left(1,\gamma\right)$.

Owing to the discrete nature and infinite dimensionality of its draws, the DP is a highly useful prior for Bayesian mixture models. By associating different mixture components with atoms $\phi_{k}$ of the stick-breaking process, and assuming $x_{i}|\phi_{k}\overset{iid}{\sim}f\left(x_{i}|\phi_{k}\right)$ where $f\left(.\right)$ is the likelihood kernel of the mixing components, we can formulate the infinite Bayesian mixture model or Dirichlet process mixture model (DPM). Approximate methods are used for posterior inference~\cite{Neal_2000_Markov}.

\subsubsection{Hierarchical Dirichlet process mixture models\label{sub:HDP}}
While DPM is suitable for non-parametric clustering of exchangeable data in a single group, many real-world problems are more appropriately modelled as comprising multiple groups of exchangeable data. In such cases it is usually desirable
to model the observations of different groups jointly, allowing them to share their generative clusters. This idea is known as ``sharing the statistical strength'' and it is naturally obtained by hierarchical architecture in Bayesian modelling.

Consider a collection of documents. DPM models each group with an infinite number of topics. However, it is desired for multiple group-level DPMs to share their clusters. Amongst different ways of linking group-level DPMs, HDP~\cite{teh_etal_2006_hierarchical} offers an interesting
solution whereby base measures of group-level DPs are drawn from a corpus-level DP. In this way the atoms of the corpus-level DP (i.e.\ topics in our case) are shared across the documents. Formally, if $\mathbf{x}=\left\{ \mathbf{x}_{1},\ldots,\mathbf{x}_{J}\right\} $ is a document collection where $\mathbf{x}_{j}=\left\{ x_{j1},\ldots,x_{jN_{j}}\right\} $ is the $j$-th document comprising $N_{j}$ words:
\begin{align}
  G_{0}|\gamma,H & \sim & \text{DP}\left(\gamma,H\right)\\
  G_{j}|\alpha_{0},G_{0} & \sim & \text{DP}\left(\alpha_{0},G_{0}\right)\\
  \theta_{ji}|G_{j} & \sim & G_{j}\\
  x_{ji}|\theta_{ji} & \sim & F\left(.|\theta_{ji}\right)
\end{align}

\begin{figure}
  \centering
  \subfigure[]{\includegraphics[width=0.38\textwidth]{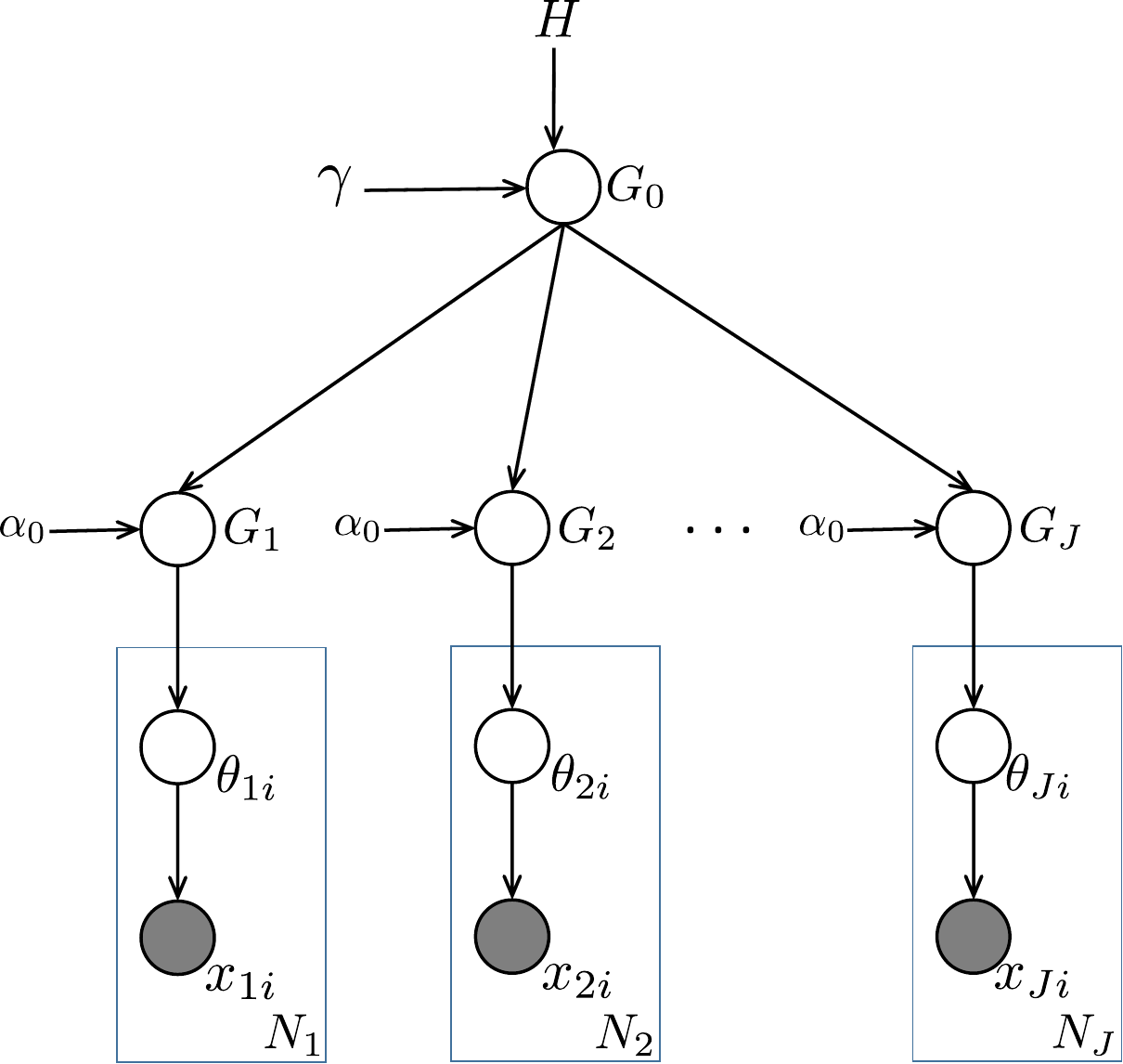}\label{fig:HDP}}
  \subfigure[]{\includegraphics[width=0.58\textwidth]{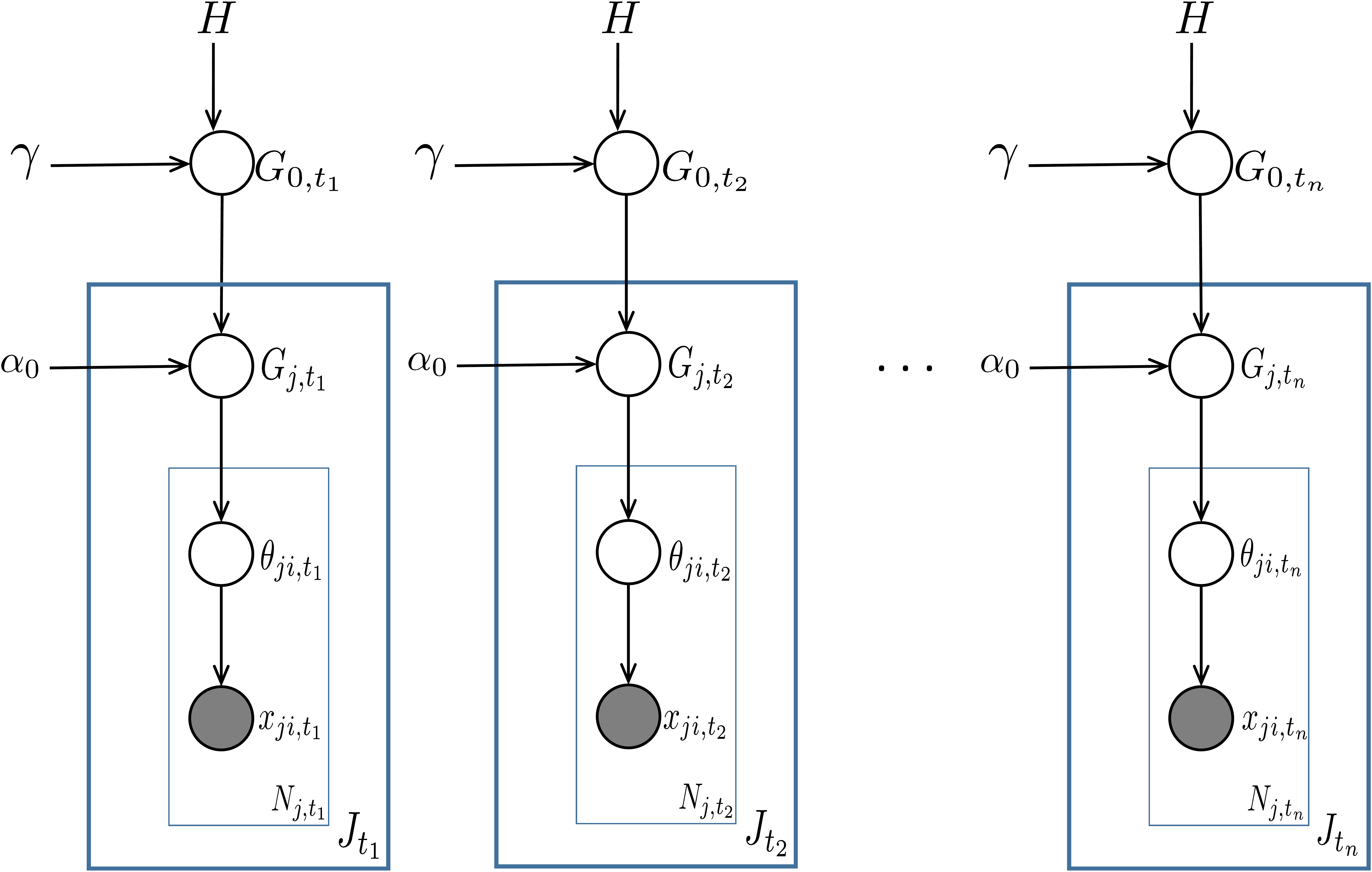}\label{f:proposedGmodel}}
  \caption{(a) Graphical model representation of HDP. Each box represents one document whose observed data (words) is shown shaded. Unshaded nodes
represent latent variables. An observed datum $x_{ji}$ is assigned to a latent mixture component parameterized by $\theta_{ji}$. $\gamma$ and $\alpha$ are the concentration parameters and $H$ is the corpus-level base measure. (b) Graphical model representation of the proposed framework. The corpus is temporally divided into $t_{n}$ epochs and each epoch modelled using an HDP (outer boxes). Different epochs' HDPs share their corpus-level DP and hyperparameters.}
\end{figure}

This is illustrated schematically in Figure~\ref{fig:HDP}. Since $G_{j}$ is drawn from a DP with base measure $G_{0}$, it takes the same support as $G_{0}$. Also the parameters of the group-level mixture components, $\theta_{ji}$, share their values with the corpus-level DP support on $\left\{ \phi_{1},\phi_{2},\ldots\right\} $. Therefore $G_{j}$ can be equivalently expressed using the stick-breaking process
as $G_{j}=\sum_{k=1}^{\infty}\pi_{jk}\delta_{\phi_{k}}$ where $\boldsymbol{\pi}_{j}|\alpha_{0},\gamma\sim\text{DP}\left(\alpha_{0},\gamma\right)$~\cite{teh2006collapsed}. The posterior for $\theta_{ji}$ has been shown to follow a Chinese restaurant franchise process which can be used to develop inference algorithms based on Gibbs sampling~\cite{teh_etal_2006_hierarchical}.

\subsection{Modelling topic evolution over time\label{ss:contrib}}
Hitherto, the discussion in this section focused on the modelling of static document corpora. We now show how our work builds on top of these ideas in the existing literature and in particular how the described HDP-based model can be applied to the analysis of temporal topic changes in a \emph{longitudinal} data corpus.

Like some of the previous work in this area we begin by discretizing time and dividing the literature corpus by time into \emph{epochs}. Each epoch spans a certain contiguous time period and has associated with it all documents with timestamps within this period. The duration of epochs should be sufficiently short to allow the corresponding document subset to be treated as a static collection; we shall discuss this issue in more detail in the next section.

Each epoch is then modelled separately using an HDP, with models corresponding to different epochs sharing their hyperparameters and the corpus-level base measure. Hence if $n$ is the number of epochs, we obtain $n$ sets of topics
$\boldsymbol{\phi}=\left\{ \boldsymbol{\phi}_{t_{1}},\ldots,\boldsymbol{\phi}_{t_{n}}\right\} $
where $\boldsymbol{\phi}_{t}=\left\{ \theta_{1,t},\ldots,\phi_{K_{t},t}\right\} $
is the set of topics that describe epoch $t$, and $K_{t}$ their number (which is inferred automatically, as described previously). This is illustrated in Figure~\ref{f:proposedGmodel}. In the next section we describe how given an inter-topic similarity measure the evolution of different topics across epochs can be tracked.

\subsubsection{Measuring topic similarity\label{ss:similarity}}
Our goal now is to track changes in the topical structure of a data corpus over time. The simplest changes of interest include the emergence of new topics, and the disappearance of others. More subtly, we are also interested in how a specific topic changes, that is, how it evolves over time in terms of the contributions of different words it comprises. Lastly, our aim is to be able to extract and model complex structural changes of the underlying topic content which result from the interaction of older topics. Specifically, topics, which can be thought of as collections of memes, can merge to form new topics or indeed split into more nuanced memetic collections. This information can provide valuable insight into the refinement of ideas and findings in the scientific community, effected by new research and accumulating evidence.

The key idea behind our approach stems from the observation that while topics may change significantly over time, providing that the duration of epochs is chosen appropriately, the change between successive epochs is limited. Therefore we infer the continuity of a topic in one epoch by relating it to all topics in the immediately subsequent epoch which are sufficiently similar to it under a suitable similarity measure. This can be seen to lead naturally to a similarity graph representation whose nodes correspond to topics and whose edges link those topics in two epochs which are related. Formally, the weight of the directed edge that links $\phi_{j,t}$, the $j$-th topic in epoch $t$, and $\phi_{k,t+1}$ is set equal to $\rho\left(\phi_{j,t},\phi_{k,t+1}\right)$ where $\rho$ is a similarity measure. Given that in our HDP-based model each topic is represented by a probability distribution, similarity measures such the Bhattacharyya coefficient~\cite{Bhat1943}, the Jenson-Shannon or Kullback-Leibler divergences~\cite{Lin1991,AranCipo2006e}, and the Hellinger~\cite{Hell1909} or resistor-average distances~\cite{RieuAran2015}, for example, are all readily adapted for use in the proposed framework as we shall demonstrate in the next section.

\begin{figure}
  \centering
    \includegraphics[width=0.6\textwidth]{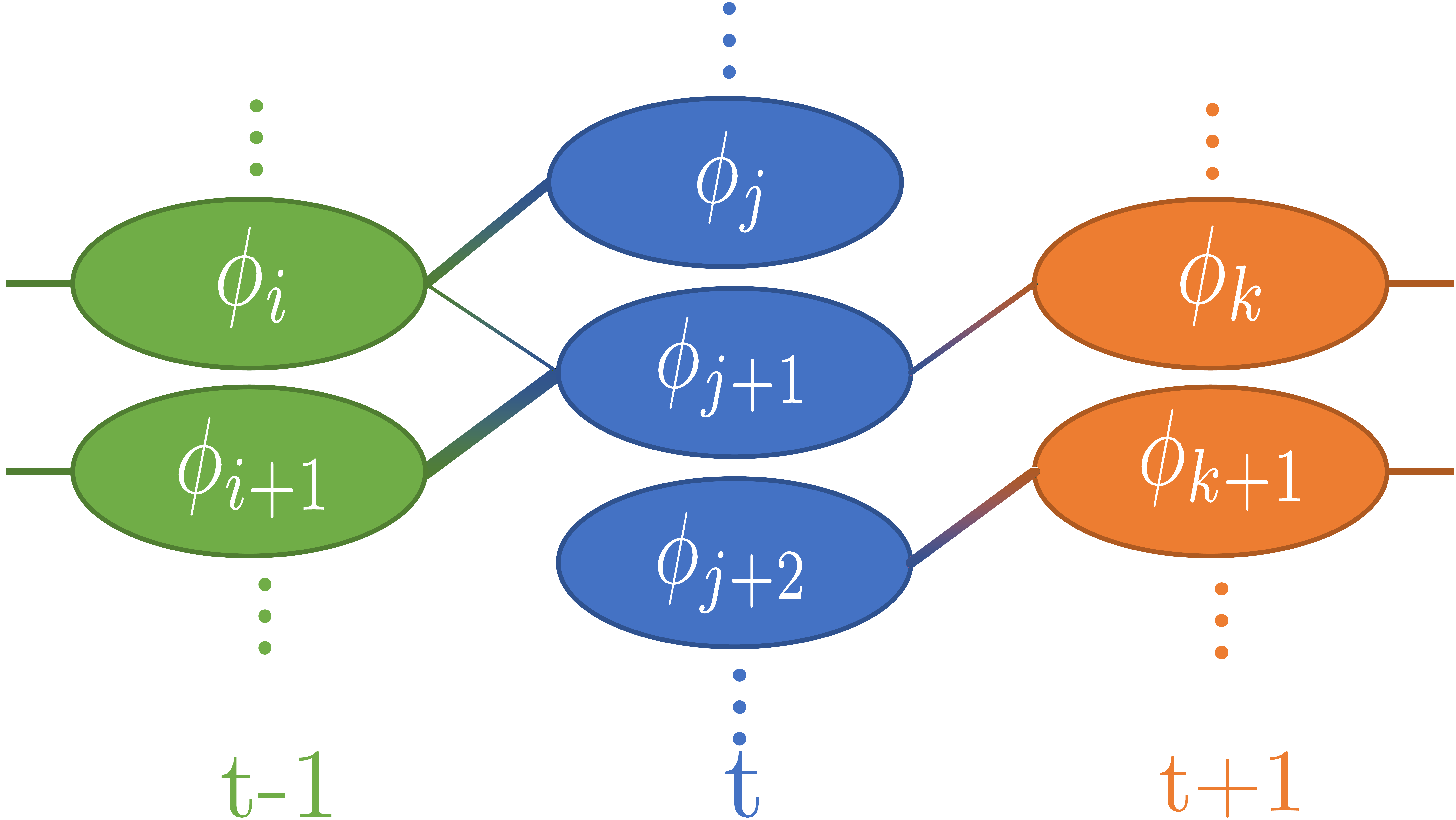}
  \caption{Conceptual illustration of the proposed similarity graph that models topic dynamics over time. A node corresponds to a topic in a specific epoch; edge weights are equal to the corresponding topic similarities.
  \label{fig:topic_similarity_graph}}
\end{figure}

A conceptual illustration of a small section of a similarity graph is shown in Figure~\ref{fig:topic_similarity_graph}. It shows three consecutive time epochs $t-1$, $t$, and $t+1$ and a selection of topics in these epochs. Graph edge weight (i.e.\ inter-topic similarity) is encoded by the thickness of the line representing the edge -- the thicker the line, the more similar the corresponding topics are. In constructing a similarity graph we use a threshold to eliminate automatically weak edges, retaining only the connections between sufficiently similar topics in adjacent epochs. This readily allows us to detect the disappearance of a particular topic, the emergence of new topics, as well as the splitting or merging of different topics as follows:
\begin{itemize}
\item \textbf{New topic emergence:}\\~If a node does not have any edges incident to it, the corresponding topic is taken as having emerged in the associated epoch; for example, in Figure~\ref{fig:topic_similarity_graph} the topic $\phi_{j+2}$ can be seen to emerge during the epoch with the timestamp $t$.

\item \textbf{Topic disappearance:}\\ If no edges originate from a node, the corresponding
topic is taken to vanish in the associated epoch; for example, in Figure~\ref{fig:topic_similarity_graph} the topic $\phi_{j}$ can be seen to disappear during the epoch with the timestamp $t$.

\item \textbf{Topic evolution:}\\ When exactly one edge originates from a node in one epoch and it is the only edge incident to a node in the following epoch, the topic is understood as having evolved in the sense that its memetic content (captured by the probability distribution over the underlying vocabulary) may have changed; for example, in Figure~\ref{fig:topic_similarity_graph} the topic $\phi_{j+2}$ evolves into the topic $\phi_{k+1}$ between the epochs with the timestamps $t$ and $t+1$.

\item \textbf{Topic splitting:}\\ If more than a single edge originates from a node, we interpret this as the corresponding topic splitting into multiple topics between the corresponding epochs and the successive epoch; for example, in Figure~\ref{fig:topic_similarity_graph} the topic $\phi_{i}$ splits into topics $\phi_{j}$ and $\phi_{j+1}$ between the epochs with the timestamps $t-1$ and $t$.

\item \textbf{Topic merging:}\\ If more than a single edge is incident to a node, the topics of the nodes from which the edges originate are understood to have interacted by merging to form a new topic; for example, in Figure~\ref{fig:topic_similarity_graph} between the epochs with the timestamps $t-1$ and $t$ the topics $\phi_{i}$ and $\phi_{i+1}$ merge to form $\phi_{j+1}$.
\end{itemize}~

Lastly, observe that just as our understanding of topics at a higher semantic level would allow, the proposed framework readily models complex structural changes which involve a topic concurrently undergoing merging and splitting. For example, the topic labelled $\phi_{j+1}$ in Figure~\ref{fig:topic_similarity_graph} is created though the merging of topic $\phi_{i+1}$ and a split offshoot of the topic $\phi_i$.

\subsubsection{Automatic temporal similarity graph construction\label{ss:construction}}
In our previous work~\cite{BeykPhunAranVenk2015,BeykAranPhunVenk2015a} the temporal similarity graph is built in two stages. In the first stage the graph is connected fully in the sense that all pairs of topics in successive epochs are connected by edges. Then, the graph is pruned using a similarity threshold $t_s$. In other words any edge corresponding to an inter-topic similarity lesser than $t_s$ is removed from the graph.

\begin{figure}
  \centering
  \includegraphics[width=0.9\textwidth]{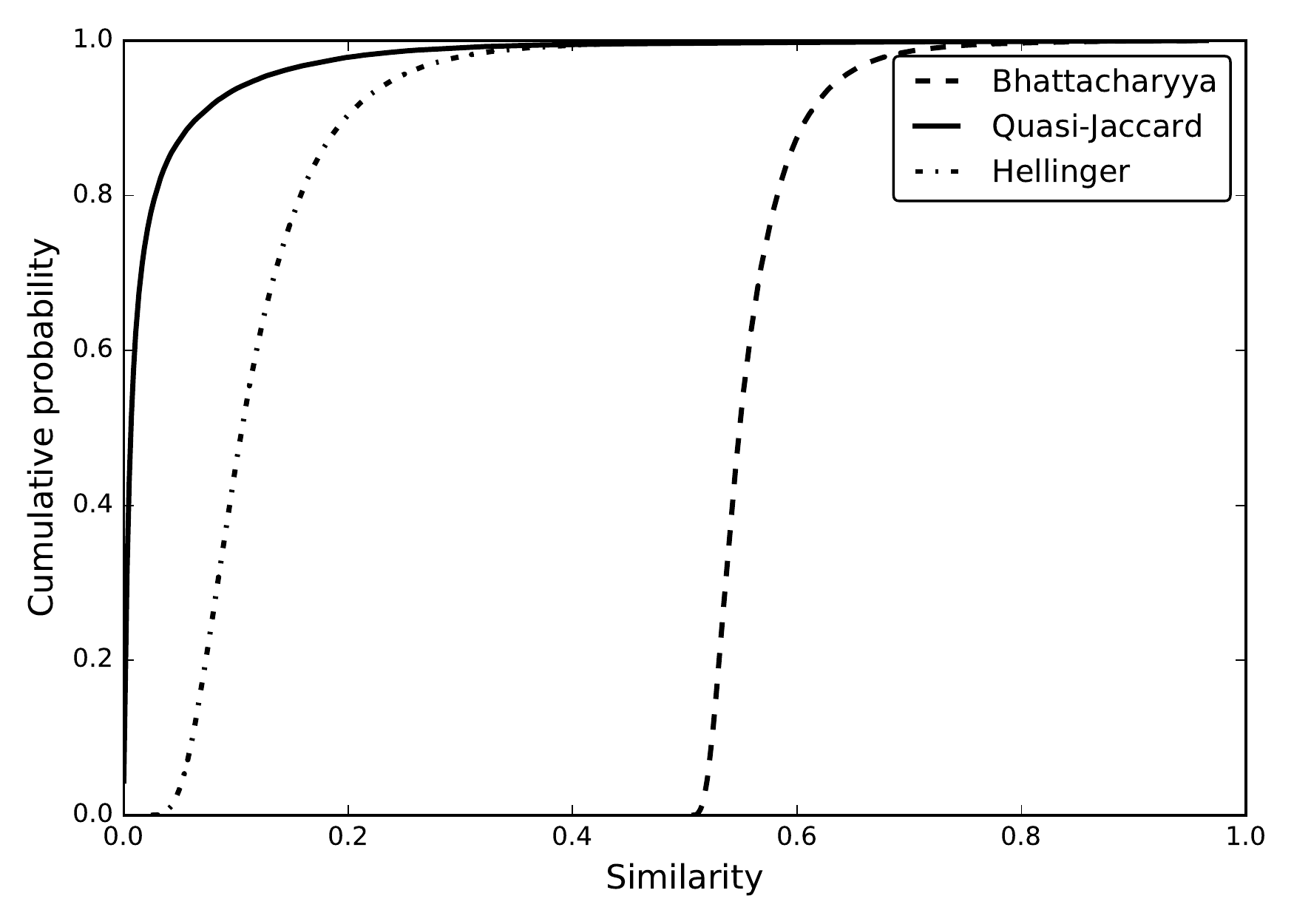}
  \caption{Empirical estimates of the cumulative density function (CDF) of inter-topic similarities corresponding to fully connected temporal similarity graph constructed over the same set of topics but using three different similarity measures (please see Section~\ref{ss:evalMeasures}). Such CDFs are used in the proposed construction of the final temporal similarity graph, that is, in the pruning process used to derive it from the precedent, fully connected graph. }
  \label{f:CDF}
\end{figure}

A major limitation of the described pruning step is that the similarity threshold $t_s$ is not readily interpretable and the framework provides little insight as to how the threshold should be chosen. In addition, considering that the threshold value inherently depends on the similarity measure which is used, it is not clear how two inter-topic similarity measures may be compared i.e.\ how to control for the threshold in the presence of a changing similarity metric which underlies it. Hence in the present paper we describe an alternative strategy which employs a more meaningful and more interpretable manner of pruning. Firstly we consider all inter-topic similarities present in the initial fully connected graph and extract the empirical 	estimate of the corresponding cumulative density function (CDF). Examples of CDFs obtained using three similarity metrics on a typical epoch in our data (please see the next section for empirical analysis) are shown for illustration in Figure~\ref{f:CDF}. The very different functional forms of the three CDFs shown reflect our previous observation that pruning on the basis of a similarity threshold is inherently dependent on the employed similarity measure, which complicates any comparative analysis. Instead of using a similarity threshold herein we prune the graph based on the operating point on the relevant CDF. In other words if $F_\rho$ is the CDF corresponding to a specific initial, fully connected graph formed using a particular similarity measure, and $\zeta \in [0, 1]$ the CDF operating point, we prune the edge between topics $\phi_{j,t}$ and $\phi_{k,t+1}$ iff $\rho(\phi_{j,t},\phi_{k,t+1}) < F^{-1}_\rho (\zeta)$.

\begin{figure}
  \centering
  \subfigure[$\zeta=0.9$]{\includegraphics[trim={0 3cm 0 4cm}, clip, width=0.85\textwidth]{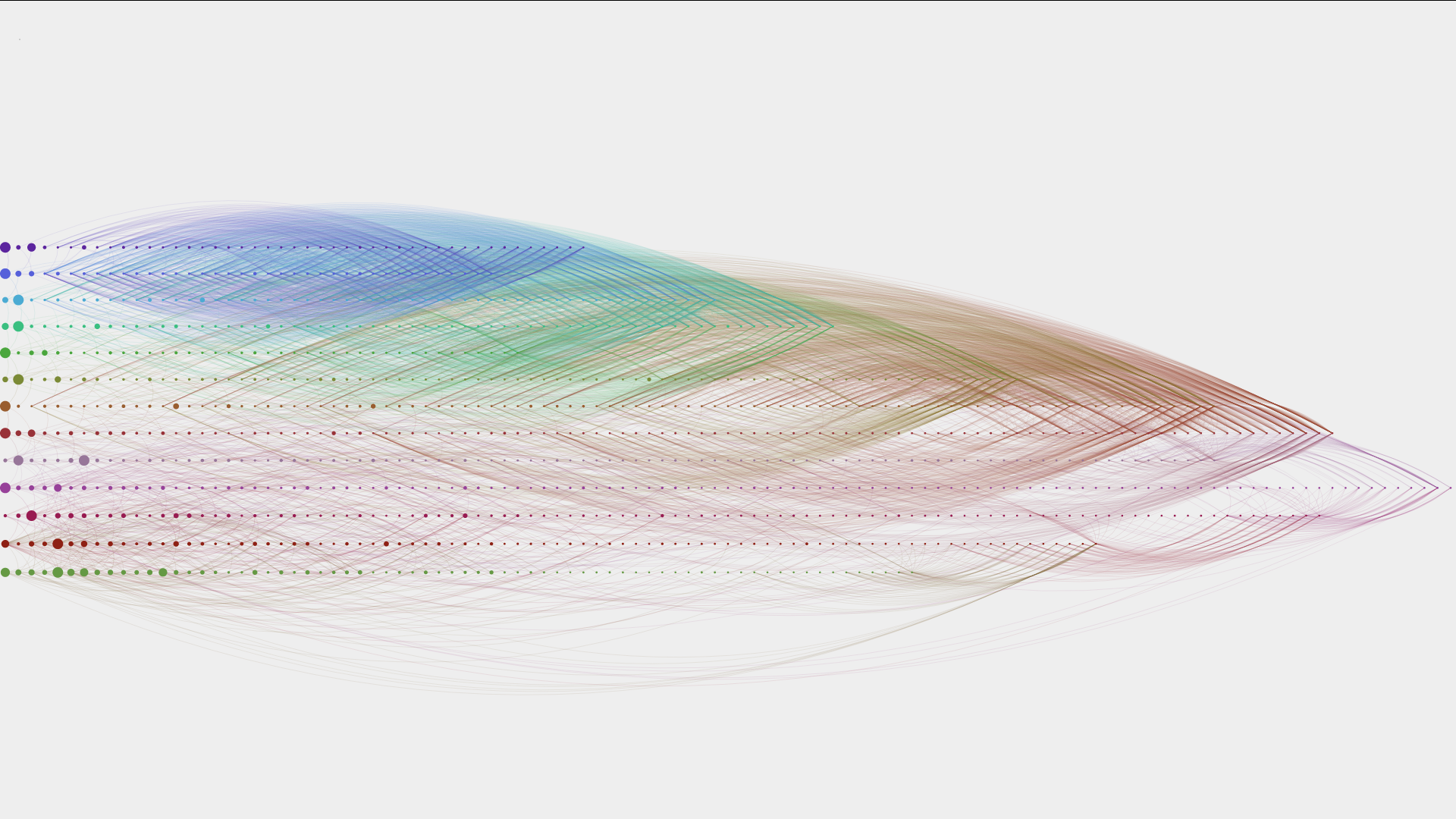}}\\[5pt]
  \subfigure[$\zeta=0.95$]{\includegraphics[trim={0 3cm 0 4cm}, clip, width=0.85\textwidth]{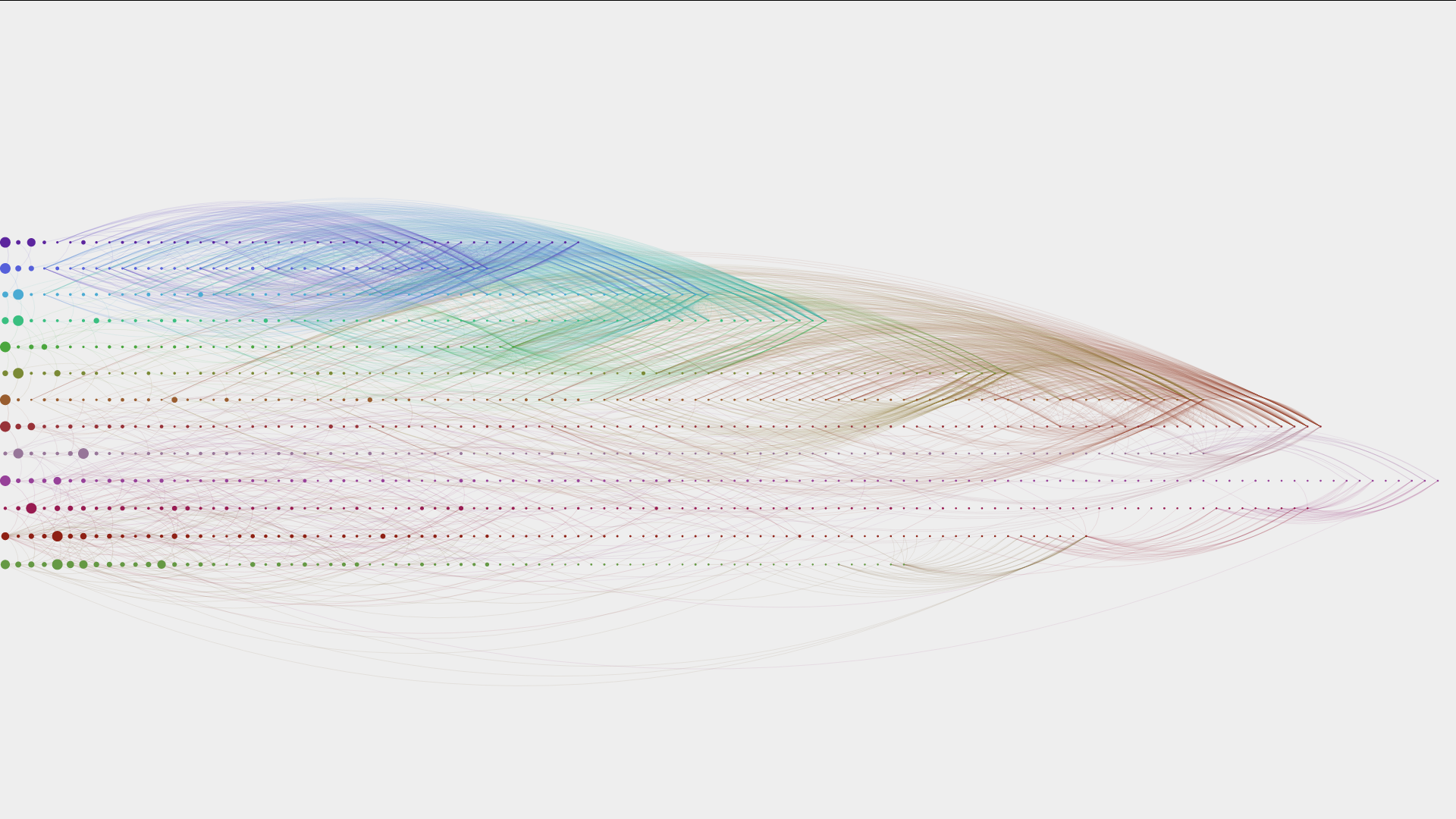}}\\[5pt]
  \subfigure[$\zeta=0.99$]{\includegraphics[trim={0 3cm 0 4cm}, clip, width=0.85\textwidth]{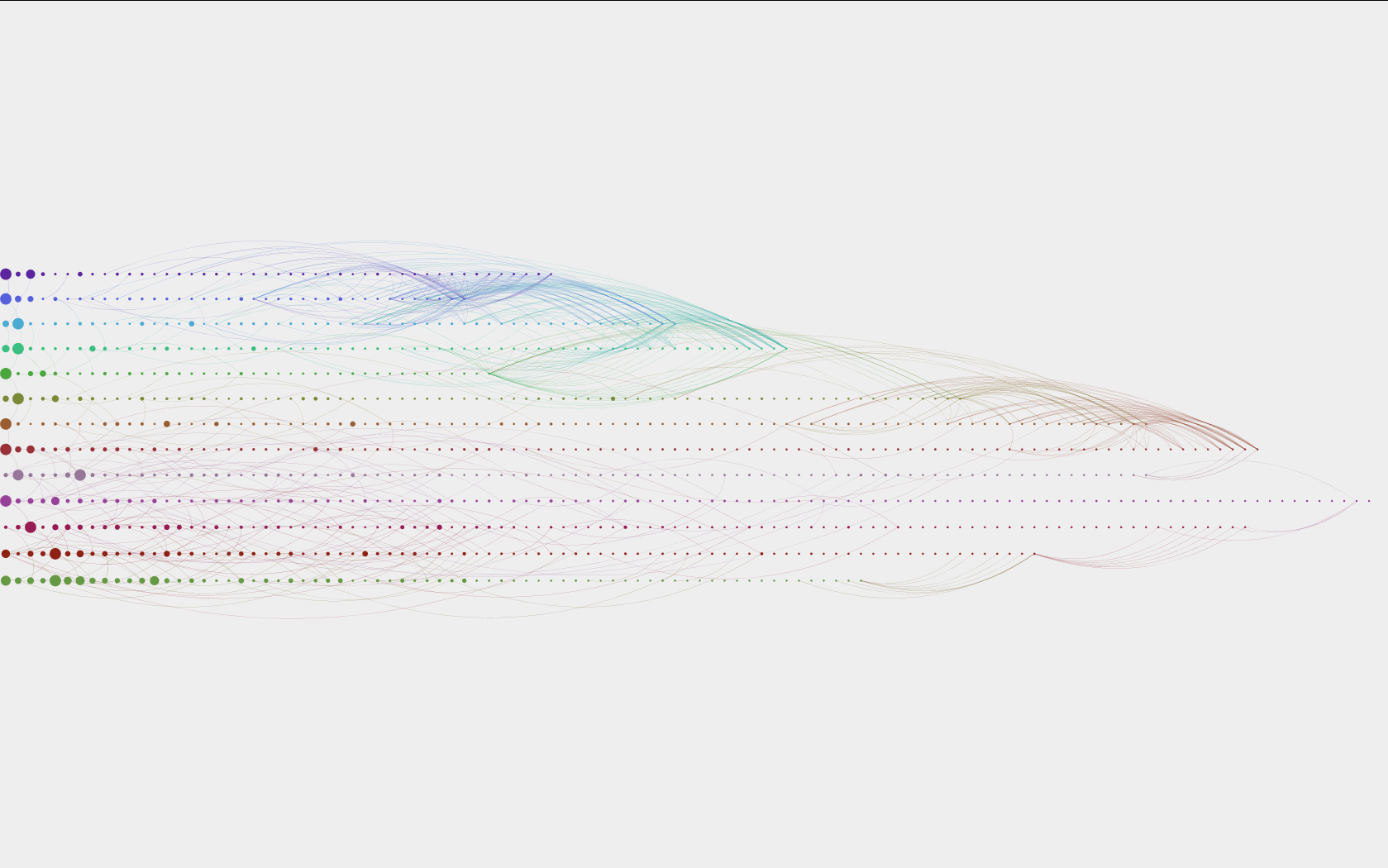}}\\[15pt]
  \caption{Examples of temporal similarity graphs. All three graphs correspond to the same data corpus and are formed over the same set of topics using a similarity measure based on the Hellinger distance (please see Section~\ref{ss:evalMeasures}) but using different CDF operating points $\zeta=0.9, 0.95, 0.99$. The horizontal axis denotes time with each column of topics (dots) corresponding to a specific epoch. The size of dots representing topics codes for their popularity in the corresponding epoch. As expected, increasing $\zeta$ results in an increase in sparsity of the final temporal graph. }
  \label{f:exampleGraphs}
\end{figure}

\section{Experimental evaluation\label{s:eval}}
Having introduced the main technical contribution of our work we now analyse the performance of the proposed framework empirically on two large real-world data sets. 

\subsection{Evaluation data}\label{ss:evalData}
In Section~\ref{s:previous} we highlighted the corpus of published peer-reviewed scientific literature as a particularly attractive data source for the application of temporal topic models. This corpus exhibits rapid growth at an accelerating pace, and it is by its very nature characterized by high fluency of ideas. These ideas can be represented well by topics in the sense in which this term is used in the present article. Hence for the evaluation of the method proposed in the preceding section we selected two highly active research areas in the realm of biomedicine.

For the sake of completeness we next briefly survey the most influential work on the analysis of biomedical texts, then outline the key reasons for our choice of the specific research areas we focus on, and then proceed with a description of the adopted experimental methodology, a presentation of the most significant results emerging from our analysis, and the associated discussion.

\subsubsection{Biomedical text mining}\label{ss:prevBiomed}
Most previous work on text-based knowledge discovery in biomedicine to date has focused on (i) the tagging of names of entities such as genes, proteins, and diseases~\cite{settles2005abner}, (ii) the discovery of relationships between different entities e.g.\ functional associations between genes~\cite{rhodes2004oncomine}, or (iii) the extraction of information pertaining to events such as gene expression or protein binding~\cite{simpson2012biomedical}.

The idea that the medical literature could be mined for new knowledge is typically attributed to Swanson~\cite{swan1986}. For example by manually examining medical literature databases he hypothesised that dietary fish oil could be beneficial for Raynaud's syndrome patients, which was later confirmed by experimental evidence. Work that followed sought to develop statistical methods which would make this process automatic. Most approaches adopted the use of term frequencies and co-occurrences using dictionaries such as Medical Subject Headings (MeSH)~\cite{Roge1963}.

Most existing work on biomedical knowledge discovery is based on what may be described as traditional data mining techniques (neural networks, support vector machines etc); comprehensive surveys can be found in~\cite{kumar2014biomedical,simpson2012biomedical}. The application of state-of-the-art Bayesian methods in this domain is scarce. Amongst the notable exceptions is the work by Blei \textit{et al.}  who showed how latent Dirichlet allocation (LDA) can be used to facilitate the process of hypothesis generation in the context of genetics~\cite{blei2006statistical}. Arnold \textit{et al.} used a similar approach to demonstrate that abstract topic space representation is effective in patient-specific case retrieval~\cite{arnold2010clinical}. In their later work they introduced a temporal model which learns topic trends and showed that the inferred topics and their temporal patterns correlate with valid clinical events and their sequences~\cite{arnold2012topic}.  Wu \textit{et al.} used LDA for gene-drug relationship ranking~\cite{wu2012ranking}.

\subsubsection{Autism spectrum disorder and the metabolic syndrome}\label{sss:asd&mets}
In this paper we evaluate our topic discovery framework on two corpora of scholarly papers. The first corpus is of abstracts of papers concerning the autism spectrum disorder, and the other of abstracts of papers related to research on the metabolic syndrome. These specific research areas are chosen for several reasons. Firstly, they concern medical issues of major practical importance -- they affect (directly or indirectly) a large number of people and impose a significant financial cost both to the society as a whole and to those affected. Secondly, the understanding of mechanisms underlying both conditions have proven to pose a significant intellectual challenge. Consequently, the dominant ideas regarding the underlying causative mechanisms, modes of treatment and their efficacy etc.\ are continually changing, experiencing both refinement as well as more abrupt paradigm shifts. These aspects make the chosen areas of research highly suitable for the evaluation of the framework described in the present work.

\paragraph{The autism spectrum disorder}
Autism spectrum disorder is a life-long neurodevelopmental disorder with poorly understood causes on the one hand, and a wide range of potential treatments supported by little evidence on the other. The disorder is characterized by severe impairments in social interaction, communication, and in some cases cognitive abilities, and typically begins in infancy or at the very latest by the age of three. ASD is recognized as comprising an aetiologically and clinically heterogeneous group of conditions whose diagnosis remains to be based solely on the complex behavioural phenotype~\cite{Mile2011}. According to the definition in the latest version (5th edition) of the Diagnostic and Statistical Manual of Mental Disorders, the autism spectrum disorder includes disorders which were previously diagnosed with more specificity as autism, Asperger syndrome, Rett syndrome, childhood disintegrative disorder, and `pervasive developmental disorder not otherwise specified'~\cite{DSM5}. Current evidence suggests that approximately 0.5-0.6\% of the population is afflicted by ASD though the actual diagnosis rate is on the increase due to the broadening diagnostic criteria~\cite{BaxtBrugErskSche+2015}. The condition is usually detected in early childhood when an abnormal lack of social reciprocity is observed.

Although the last few decades have seen significant progress in the study of ASD, the still relatively poorly understood aetiology of the condition, its phenotypical heterogeneity~\cite{LevyMandSchu2009}, and stigma associated with mental conditions~\cite{Gray1993}, have all contributed to the penetration of beliefs, and behavioural and educational interventions which are often questionable~\cite{WarrMcPhSathFoss2011} and poorly supported by evidence (e.g.\ gluten-free and casein-free diets, and cognitive behavioural therapy~\cite{DaniWood2013}), and sometimes outright in conflict with science~\cite{TremBalaRoss2005}. For example a recent review of early intensive behavioural and developmental interventions for young children with ASD found 1 existing study as being of good quality, 10 as fair quality, and 23 as poor quality~\cite{WarrMcPhSathFoss2011}. From the public policy point of view, understanding the practices and beliefs of parents and carers of ASD-affected individuals is crucial, yet often lacking~\cite{HarrRoseGarnPatr2006}.

\paragraph{Metabolic syndrome}
Much like ASD, metabolic syndrome (also known as insulin resistance syndrome and syndrome X) does not describe a single disorder but rather a cluster of interconnected health risk factors~\cite{GrunBrewCleeSmit+2004}. Specifically, the diagnostic criterion is the presence of at least three of the following: visceral obesity, arterial hypertension, hyperglycaemia, hypertriglyceridemia, and hypoalphalipoproteinemia~\cite{GrunBrewCleeSmit+2004}. MetS is recognized as a major and escalating public health challenge, chiefly in the developed world, and is thought to be caused in part by excess energy intake, and decreased energy output due to an increasingly sedentary lifestyle~\cite{LakkLaakLakkMann+2003}. Metabolic syndrome is associated with an increased risk of numerous diseases and particularly notably with the development of cardiovascular disease and type 2 diabetes mellitus~\cite{AlbeEckeGrunZimm+2009}. Approximately one third of adults in the USA suffer from MetS~\cite{FordGileDiet2002}, with the prevalence of the syndrome increasing with age~\cite{FordGileDiet2002}.

\subsubsection{Data collection\label{ss:data}}
To the best of our knowledge there are no publicly available corpora of ASD or MetS related medical literature. Hence we collected them ourselves. These are now publicly available and can be downloaded from: \texttt{https://oa7.host.cs.st-andrews.ac.uk/}. The data sets, their collection, and preparation are described next.

\subsubsection{Raw data collection}\label{sss:rawData}
We used the PubMed interface to access the US National Library of Medicine and retrieve abstracts and references of life science and biomedical scholarly articles. We assumed that a paper is related to ASD or MetS respectively if the terms ``autism'' or ``metabolic syndrome'' are present in its title or abstract, and collected only papers written in English. The earliest publications fitting our criteria are by Kanner~\cite{kanner1946irrelevant} on ASD and by Berardinelli~\textit{et al.}~\cite{berardinelli1953new} on MetS. We collected all matching publications up to the final one indexed by PubMed on 10th May 2015, yielding a corpus of 22,508 on ASD and 31,706 publications on MetS. We used the abstract text to evaluate our method.

\subsubsection{Data pre-processing\label{sss:preprocessing}}
Data collected in the manner described in the previous section comprises abstracts as freeform text. To prepare it for the type of analysis described in Section~\ref{s:proposed} we perform a series of `pre-processing' steps. The goal is to remove words which are largely uninformative in any context, reduce dispersal of semantically equivalent terms, and thereafter select terms which are included in the vocabulary over which topics are learnt.

We firstly applied soft lemmatization using the WordNet$^\circledR$ lexicon~\cite{Mill1995} to normalize for word inflections. No stemming was performed to avoid semantic distortion often effected by heuristic rules used by stemming algorithms. After lemmatization and the removal of so-called stop-words, we obtained approximately 2.2 and 3.8 million terms in the entire corpus when repetitions are counted, and 37,626 and 46,114 unique terms, for the ASD and MetS corpora respectively. Constructing the vocabulary for our method by selecting the most frequent terms which explain 90\% of the energy in a specific corpus resulted in ASD and MetS vocabularies containing 3,417 and 2,839 terms respectively.

\subsection{Inter-topic similarity measures}\label{ss:evalMeasures}
Recall that in this work topics are probability distributions over a fixed vocabulary of terms. Thus the inter-topic similarity measure used to construct our temporal graph are thus similarity measures between probability distributions. Considering that the vocabulary of terms is fixed we represent each topic, say $p$, using a fixed length vector:
\begin{align}
  p = \left[ p_1, p_2, \ldots, p_{n_v}  \right]^T
\end{align}
where $n_v$ is the number of terms in the vocabulary.

For the experiments in this paper we adopted three well known measures for quantifying the similarity (or equivalently, dissimilarity) between probability distributions representing extracted topics. The first of these is the well known Hellinger distance~\cite{Hell1909}. For two discrete probability distributions, e.g.\ $p$ and $q$ representing two topics, it is defined as follows: 
\begin{align}
  H(p,q) = \frac{1} {\sqrt{2}} \sqrt{ \sum_{i=1}^{n_v}  (\sqrt{p_i} - \sqrt{q_i})^2 }.
  \label{e:hellinger}
\end{align}
It can be readily seen that $H()$ is symmetric and that it takes on a value between 0 and 1, with 0 signifying the greatest degree of similarity between $p$ and $q$ (in this case $p=q$) and 1 the least (in this case $p_i > 0 \implies q_i=0 ~~\land~~ q_i > 0 \implies p_i=0$).

The second similarity measure we evaluate is the Bhattacharyya coefficient defined as~\cite{Bhat1943}:
\begin{align}
  B(p,q) = \sum_{i=1}^{n_v}  \sqrt{p_i  q_i}.
  \label{e:bhattacharyya}
\end{align}
As the Hellinger distance, the Bhattacharyya coefficient is symmetric and takes on a value between 0 and 1. However note that in this case it is the maximum value of 1 which is attained when there is the greatest degree of similarity between $p$ and $q$ (i.e.\ $p=q$) and 0 the least (as before in this case $p_i > 0 \implies q_i=0 ~~\land~~ q_i > 0 \implies p_i=0$). It is straightforward to demonstrate that the Bhattacharyya coefficient is related to the Hellinger distance as follows:
\begin{align}
  H(p,q) = \sqrt{ 1 - B(p,q) }.
  \label{e:HellBhatt}
\end{align}

Lastly, due to its widespread use we also compare the performance of the proposed algorithm using the similarity measure often erroneously referred to as Tanimoto similarity~\cite{Lipk1999} (or the Jaccard similarity~\cite{FligVerdBlow2002}), defined as:
\begin{align}
  T(p,q) = \frac{ \sum_{i=1}^{n_v} p_i q_i }{\sum_{i=1}^{n}{p_i}^2+\sum_{i=1}^{n_v} {q_i}^2-\sum_{i=1}^{n_v}p_i q_i}.
  \label{e:tanimoto}
\end{align}
As the Bhattacharyya coefficient, this measure is symmetric and takes on a value between 0 and 1, with the maximum value of 1 being attained when $p$ and $q$ are equal, and 0 when there is no overlap between them (i.e.\ $p_i > 0 \implies q_i=0 ~~\land~~ q_i > 0 \implies p_i=0$). In an effort to avoid perpetuating the aforementioned misnomer on the one hand while retaining some nomenclatural continuity with the existing literature, we will refer to this measure as quasi-Jaccard similarity.

\subsection{Experiments and results}\label{ss:exp}
In this section we conduct experiments on the two corpora of scholarly literature described in Section~\ref{ss:data}, and report and discuss our results both using quantitative findings and representative qualitative examples.

\subsubsection{Quantitative comparison}\label{sss:quant}
We started our evaluation with an experiment examining quantitative differences effected by changing different flexible parameters of the proposed method.  In particular, our aim was to see how the evolving topic structure extracted by our algorithm is affected by different choices of inter-topic similarity measures (introduced in described in Section~\ref{ss:similarity} and used to construct the temporal similarity graph as described in Section~\ref{ss:construction}), different pruning thresholds used to infer complex structural changes, lengths of epochs used to discretize the timespan of a data corpus, and the overlap between successive epochs. As explained in the previous section, we compared three inter-topic similarity measures based on the Hellinger distance, the Bhattacharyya coefficient, and what we term the quasi-Jaccard similarity. Different combinations of epoch lengths and successive epoch overlaps are summarized in Table~\ref{t:paramsASD} and Table~\ref{t:paramsMS} for respectively the ASD and the metabolic syndrome data sets. Notice that due to the different timespans of the two data sets, different epoch lengths were used in the corresponding experiments. Similarly, epoch overlaps were adjusted (to the closest year) so as to be comparable on a relative basis so that, for example, the maximum overlap examined was approximately half of the duration of an epoch ($10:5$ vs.\ $6:3$). At the other extreme we also performed experiments using no epoch overlap, as well as an additional setting per experiment with an overlap of approximately 25\% of epoch length ($10:2$ and $5:1$ in Table~\ref{t:paramsASD}, and $6:1$ in Table~\ref{t:paramsMS}) .

\begin{table}
  \centering
  \setlength{\tabcolsep}{0.5cm}
  \caption{Different combinations of values of the free parameters of the proposed algorithm used for the experiments on our autism spectrum disorder abstracts data set.  }
  \vspace{8pt}
  \begin{tabular}{lcccccc}
    \Hline
    Setting \# & 1 & 2 & 3 & 4 & 5 & 6\\
    \fline
    Epoch length (years)  & 10 & 10 & 10 & 5 & 5 & 5\\
    Epoch overlap (years) & 5   & 2   &   0 & 2 & 1 & 0\\
    \Hline
  \end{tabular}
  \label{t:paramsASD}
  \vspace{0pt}
\end{table}

\begin{table}
  \centering
  \setlength{\tabcolsep}{0.5cm}
  \caption{Different combinations of values for the free parameters of the proposed algorithm used in our comparative evaluation experiments on the metabolic syndrome abstracts data set.  }
  \vspace{8pt}
  \begin{tabular}{l|ccccc}
    \Hline
    Setting \# & 1 & 2 & 3 & 4 & 5 \\
    \fline
    Epoch length (years)   & 6 & 6 & 6 & 3 & 3 \\
    Epoch overlap (years) & 3 & 1 & 0 & 1 & 0\\
    \Hline
  \end{tabular}
  \label{t:paramsMS}
\end{table}

\begin{figure}
  \centering
  \subfigure[Hellinger, 10 year epochs, 5 year overlap, ASD abstracts]{\includegraphics[clip, trim={0 0.4cm 0 0.1cm}, width=0.825\textwidth]{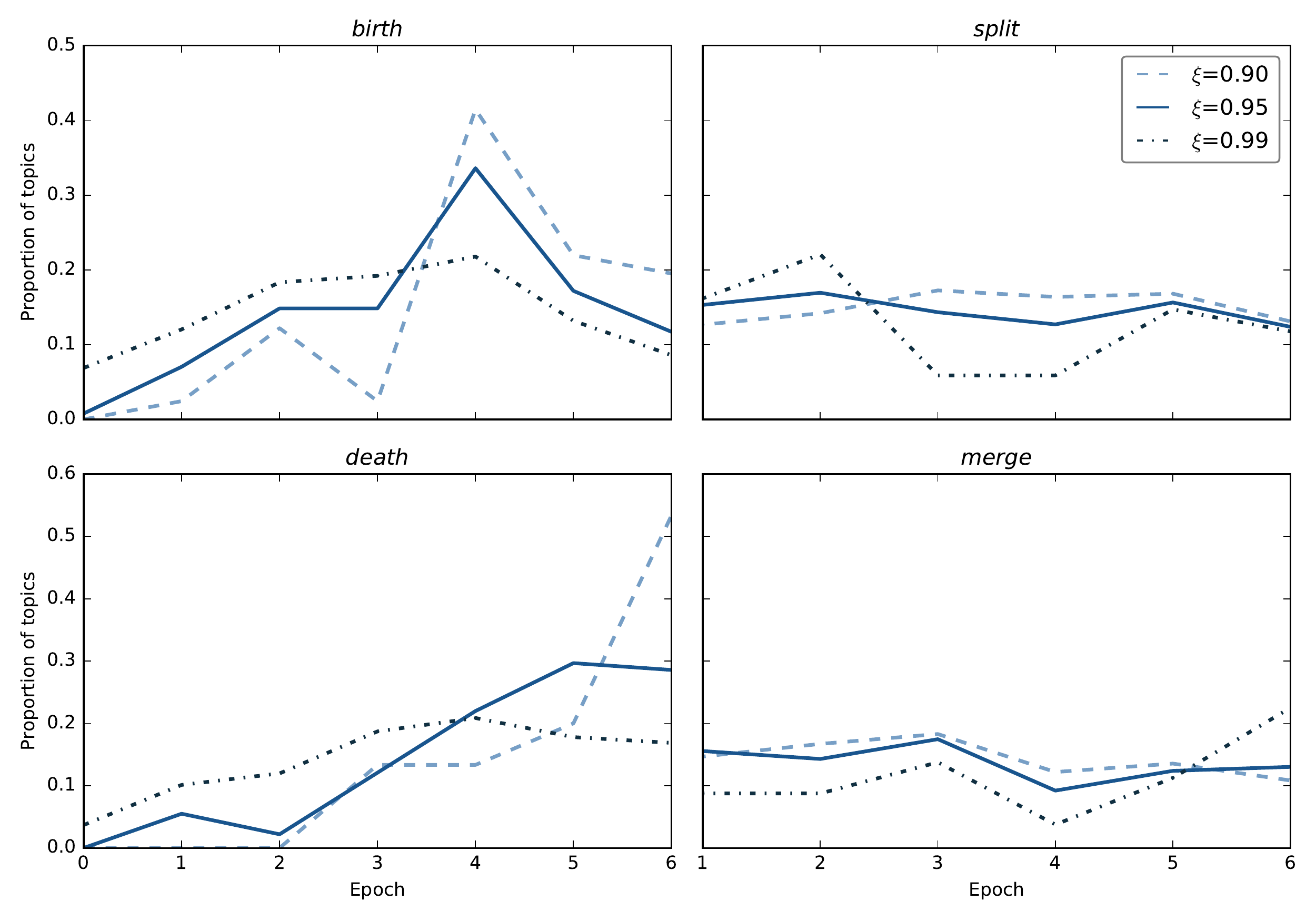}}
  \subfigure[Quasi-Jaccard, 10 year epochs, 5 year overlap, ASD abstracts]{\includegraphics[clip, trim={0 0.4cm 0 0.1cm}, width=0.825\textwidth]{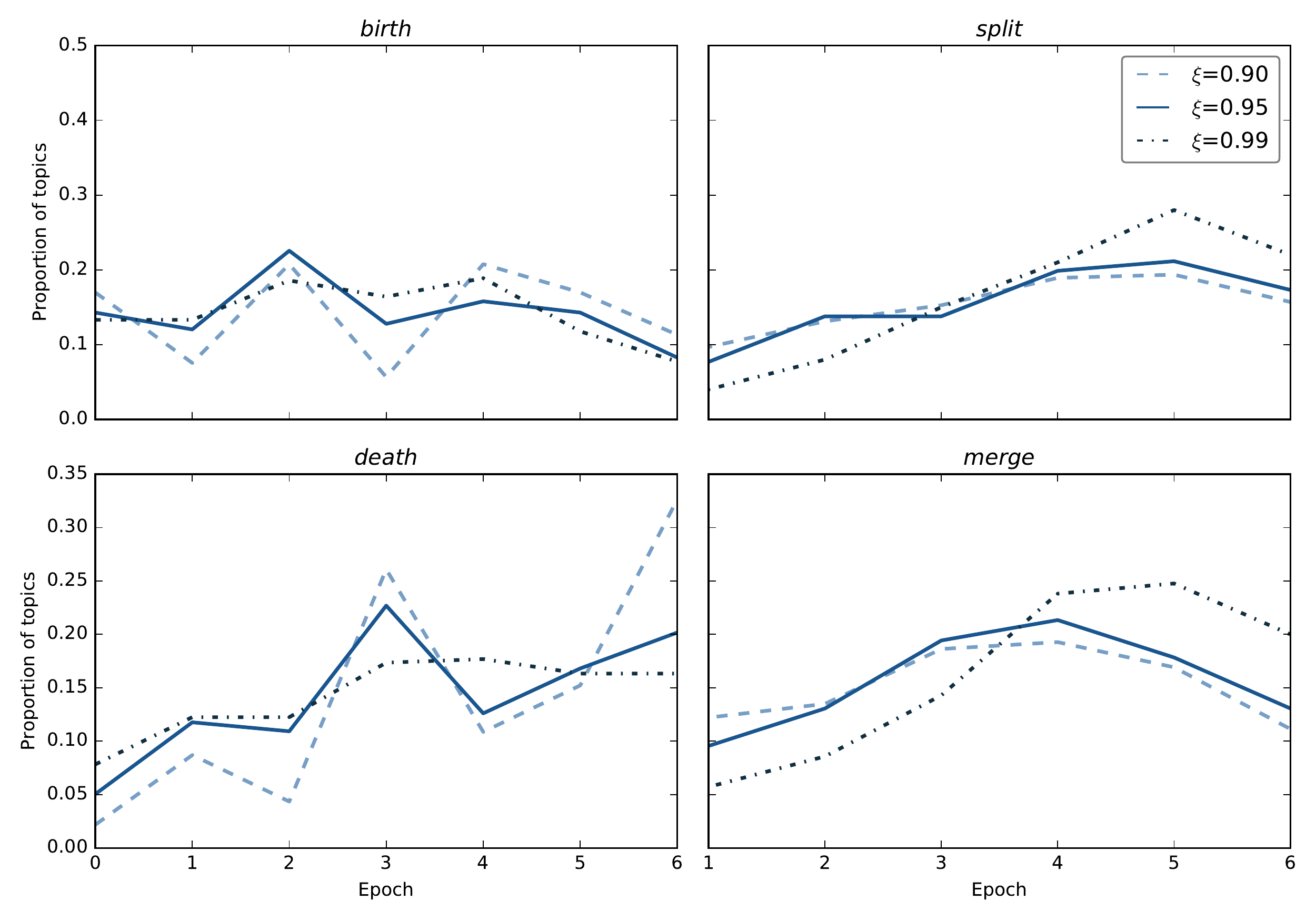}}
  \subfigure[Bhattacharyya, 10 year epochs, 5 year overlap, ASD abstracts]{\includegraphics[clip, trim={0 0.4cm 0 0.1cm}, width=0.825\textwidth]{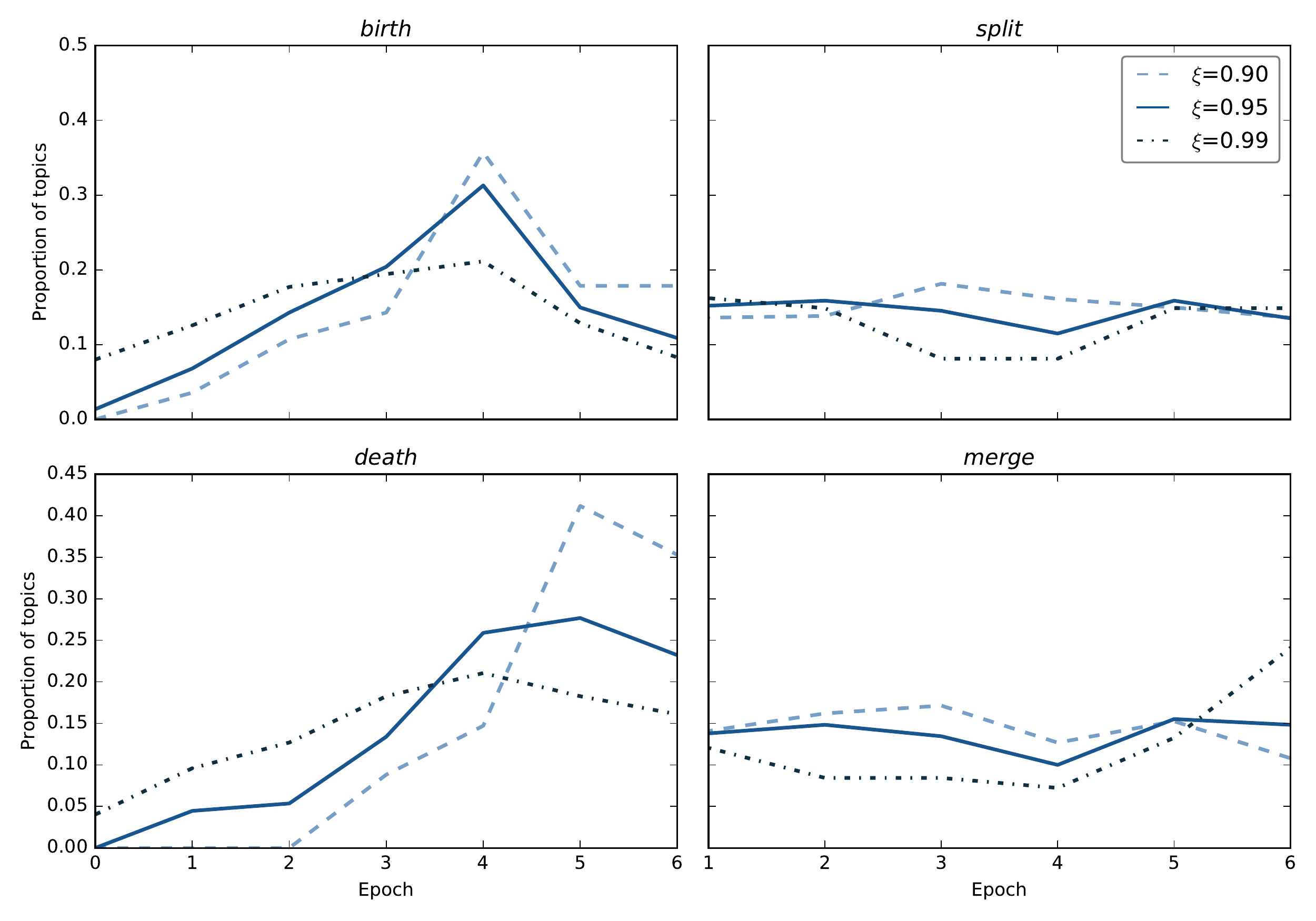}}
  \caption{Comparison of different inter-topic similarity measures using fixed values of the remaining algorithm parameters (epoch length, amount of successive epoch overlap, and cutoff operating points on the similarity CDF used to construct our temporal similarity graph.}
  \label{f:comparisonMeasures}
\end{figure}

\paragraph{Similarity measures}
To begin with, consider the results summarized in Figure~\ref{f:comparisonMeasures}. The figure comprises three sets of plots with each set corresponding to one of the three compared inter-topic similarity measures, and showing the number of topic births, deaths, merges, and splits per epoch, normalized by the total number of topics inferred in that epoch. Experiments were performed using three different operating cutoff points on the empirical cumulative density function of inter-topic similarities across the initial temporal graph -- the results are shown using lines of different styles, as per the plot legends. The epoch length and the amount of overlap of successive epochs were kept constant in all experiments.

\begin{figure}
  \centering
  \includegraphics[width=0.99\textwidth]{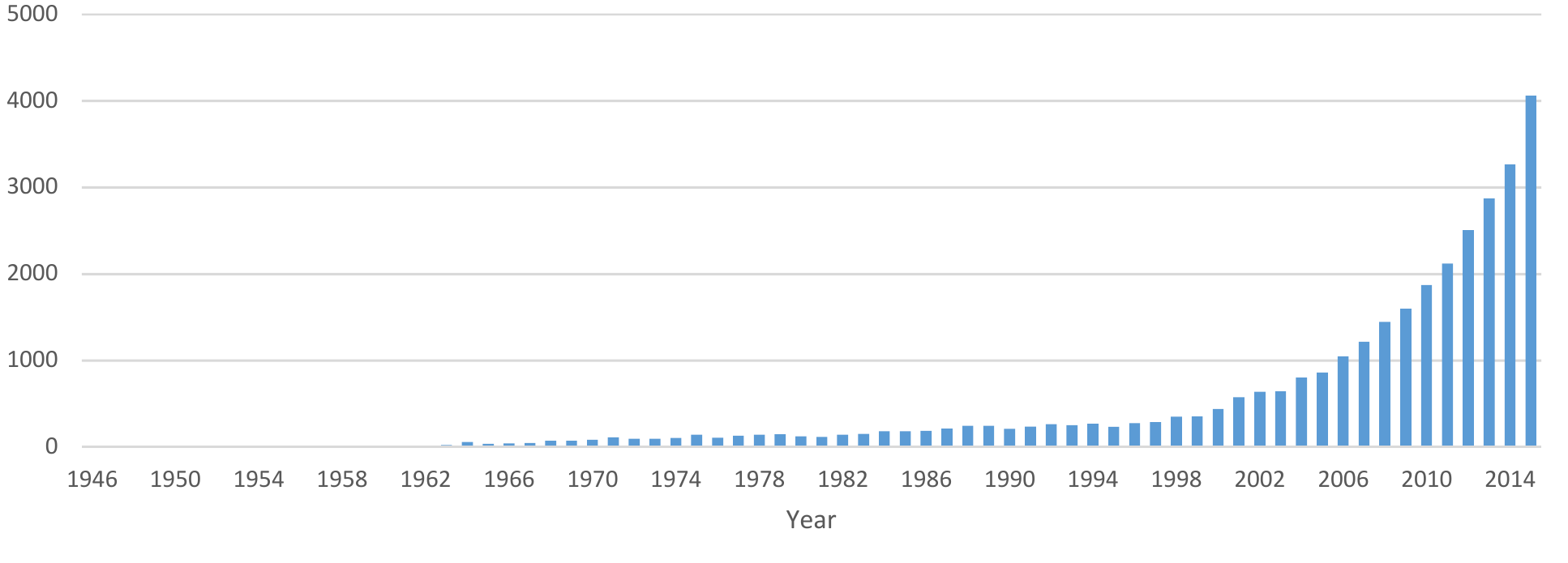}
  \caption{The rapid rise in the rate of publications concerned with ASD. Shown is the number of publications per annum indexed by PubMed and matching our criterion for the inclusion in our ASD abstracts data set (please see Section~\ref{sss:rawData}). }
  \label{f:litASD}
\end{figure}

Comparing the corresponding plots across different similarity measures, it can be readily observed that the results obtained using the Hellinger distance and the Bhattacharyya coefficient are very similar. This is unsurprising considering the close relatedness of the two measures, as highlighted previously in Section~\ref{ss:evalMeasures} and as expressed by the expression in~\eqref{e:HellBhatt}. However, at first sight the results obtained using the quasi-Jaccard index appear rather different. This raises the question which similarity measure is better (and on what grounds can such a claim be made given the fundamental lack of objective ground truth), or in a weaker form, which similarity measure should be preferred and when i.e.\ if this preference should be universal, application driven, or in some sense intimated by data itself. However a closer inspection of the plots reveals interesting and important insight that obviates this daunting question. In particular it can be noticed that while there indeed are differences in behaviour between the plots corresponding to the quasi-Jaccard index and those which correspond to the Hellinger distance and the Bhattacharyya coefficient in early epochs, eventually the three converge to approximately the same terminal behaviour. The initial differences are readily explained by the well known small sample size effect -- in the early stages of processing a longitudinal corpus inference relies on much fewer documents than later on, especially in the particular case considered here given that the amount of published research in ASD has been growing rapidly year after year as illustrated in Figure~\ref{f:litASD}. This hypothesis is further supported by the observation that the choice of different operating points on the CDF of inter-topic similarities across the initial temporal graph, used for its subsequent pruning, also has little effect in the long run -- following transient changes, the rates of different types of topic changes converge towards the same pattern.

\begin{figure}
  \centering
  \subfigure[Hellinger, 5 year epochs, 2 year overlap, ASD abstracts]{\includegraphics[clip, trim={0 0.4cm 0 0.1cm}, width=0.825\textwidth]{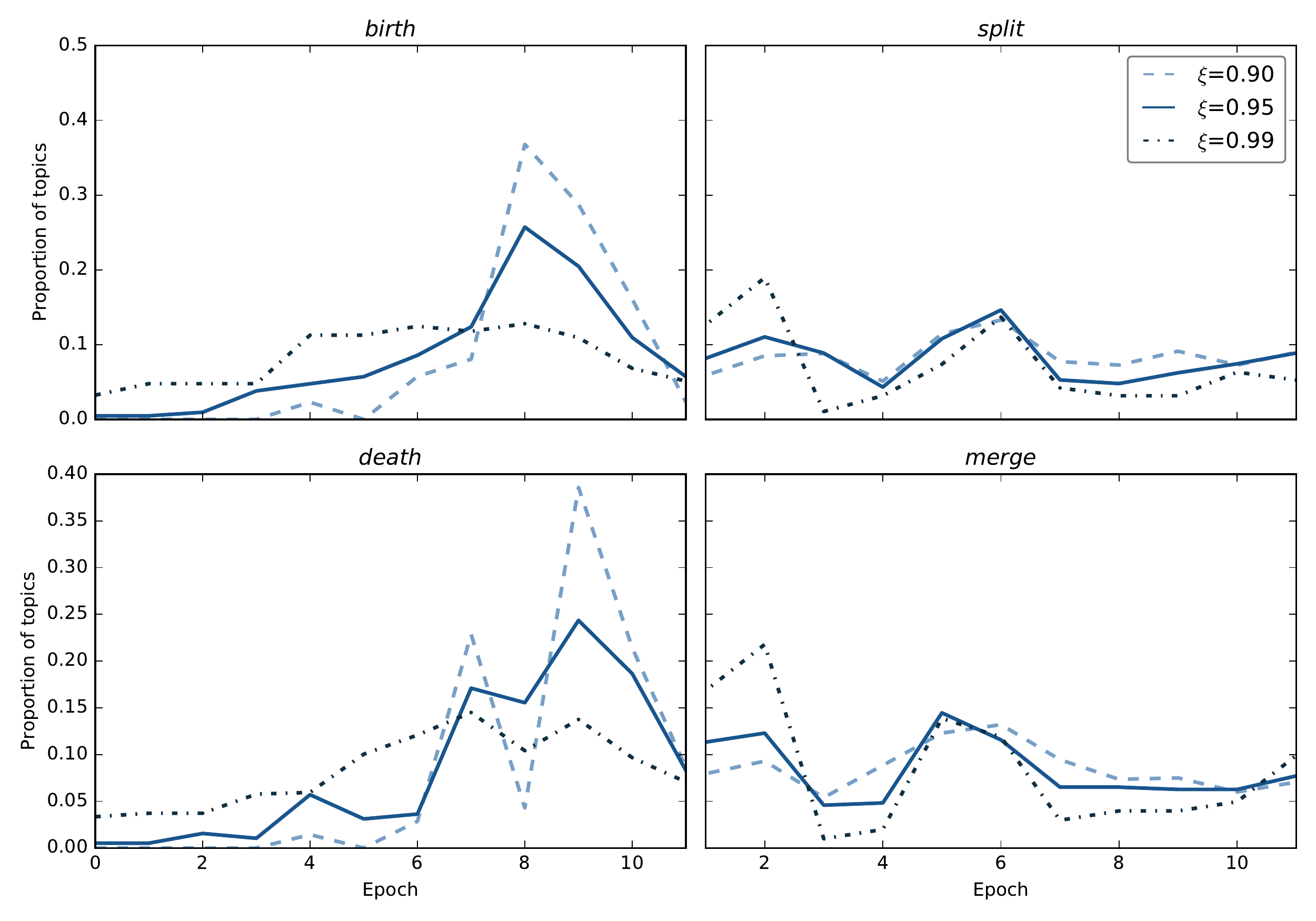}}
  \subfigure[Hellinger, 10 year epochs, 5 year overlap, ASD abstracts]{\includegraphics[clip, trim={0 0.4cm 0 0.1cm}, width=0.825\textwidth]{ASD_topic_life_per_epoch_L10O5_Hellinger_normalized.pdf}}
  \caption{The effect of the epoch length; ASD abstracts data set.}
  \label{f:epochLength1}
\end{figure}

\begin{figure}
  \centering
  \subfigure[Hellinger, 3 year epochs, 1 year overlap, MetS abstracts]{\includegraphics[clip, trim={0 0.4cm 0 0.1cm}, width=0.85\textwidth]{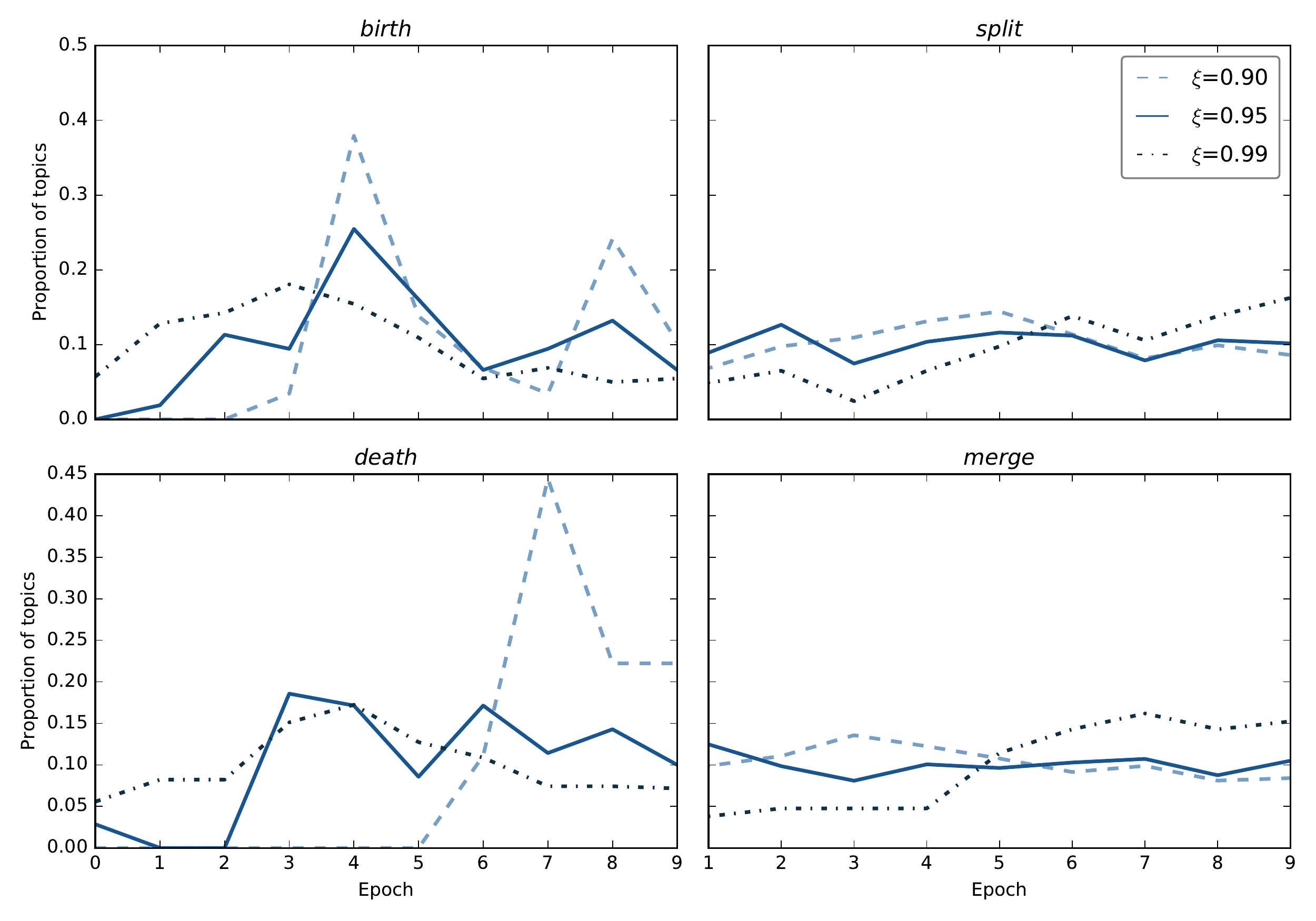}}
  \subfigure[Hellinger, 6 year epochs, 3 year overlap, MetS abstracts]{\includegraphics[clip, trim={0 0.4cm 0 0.1cm}, width=0.85\textwidth]{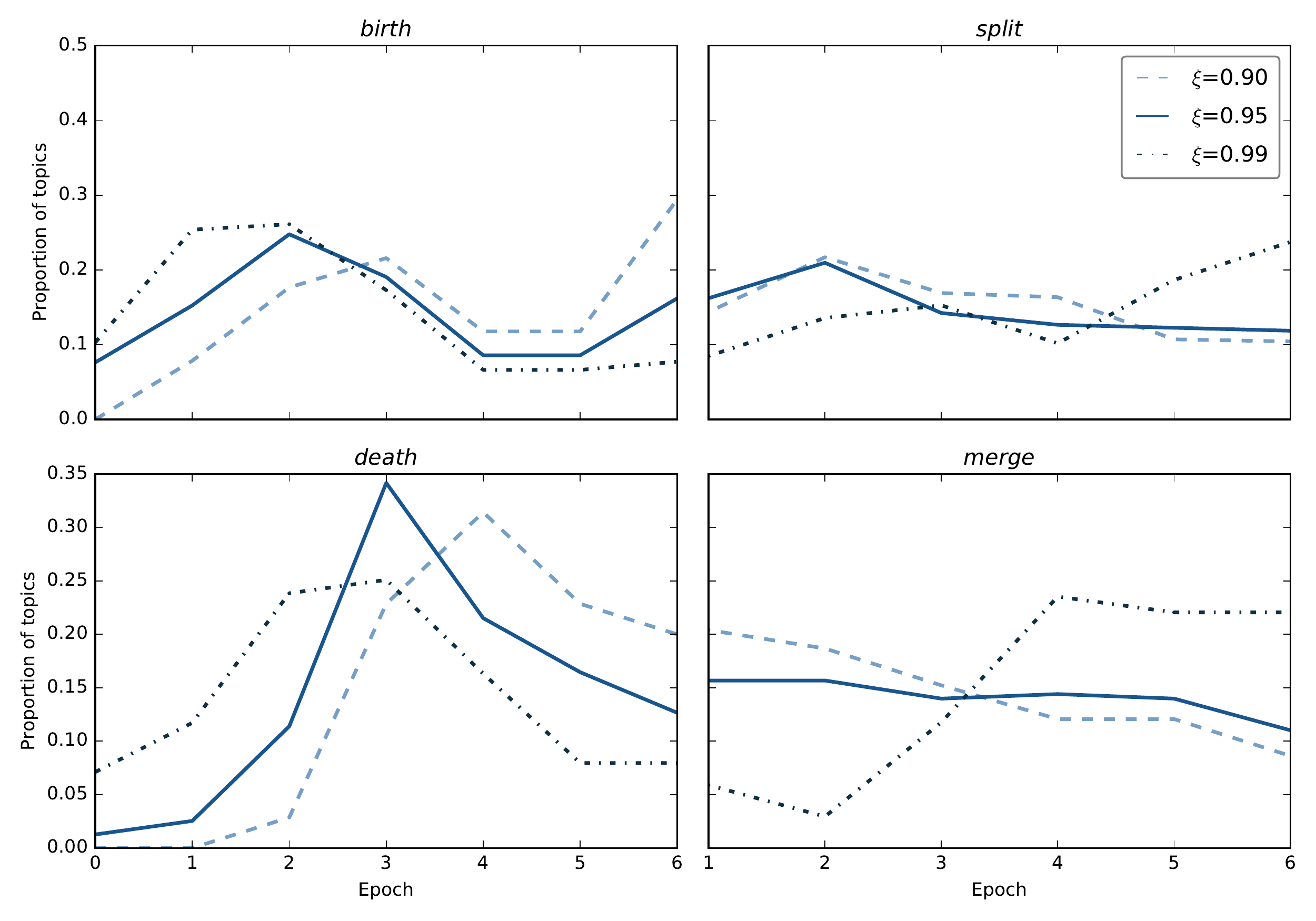}}
  \caption{The effect of the epoch length; MetS abstracts data set.}
  \label{f:epochLength2}
\end{figure}

\paragraph{Epoch length}
We next examined the effects that the choice of the epoch length has on the output of our algorithm. A representative set of results is shown in Figures~\ref{f:epochLength1} and~\ref{f:epochLength2}. Each figure comprises two sets of plots with each set corresponding to a particular set of algorithm parameters. The key parameter which was varied was the epoch length. For example, in Figure~\ref{f:epochLength1} the results obtained using the epoch length of 5 years was compared with those obtained using the epoch length of 10 years on the ASD abstracts data set. To ensure a fair comparison, successive epoch overlap was not kept constant on an absolute but rather relative basis.  Specifically, the overlap is in all cases approximately 50\% of the epoch length ($5:2$ and $10:5$ in Figure~\ref{f:epochLength1}, and $6:3$ and $3:1$ in Figure~\ref{f:epochLength2}). As before each set of plots displays the number of topic births, deaths, merges, and splits per epoch, normalized by the total number of topics inferred in that epoch, with the three different operating cutoff points on the empirical cumulative density function of inter-topic similarities across the initial temporal graph shown using lines of different styles, as per the plot legends.

\begin{figure}
  \centering
  \subfigure[Hellinger, 10 year epochs, no overlap, ASD abstracts]{\includegraphics[clip, trim={0 0.4cm 0 0.1cm}, width=0.825\textwidth]{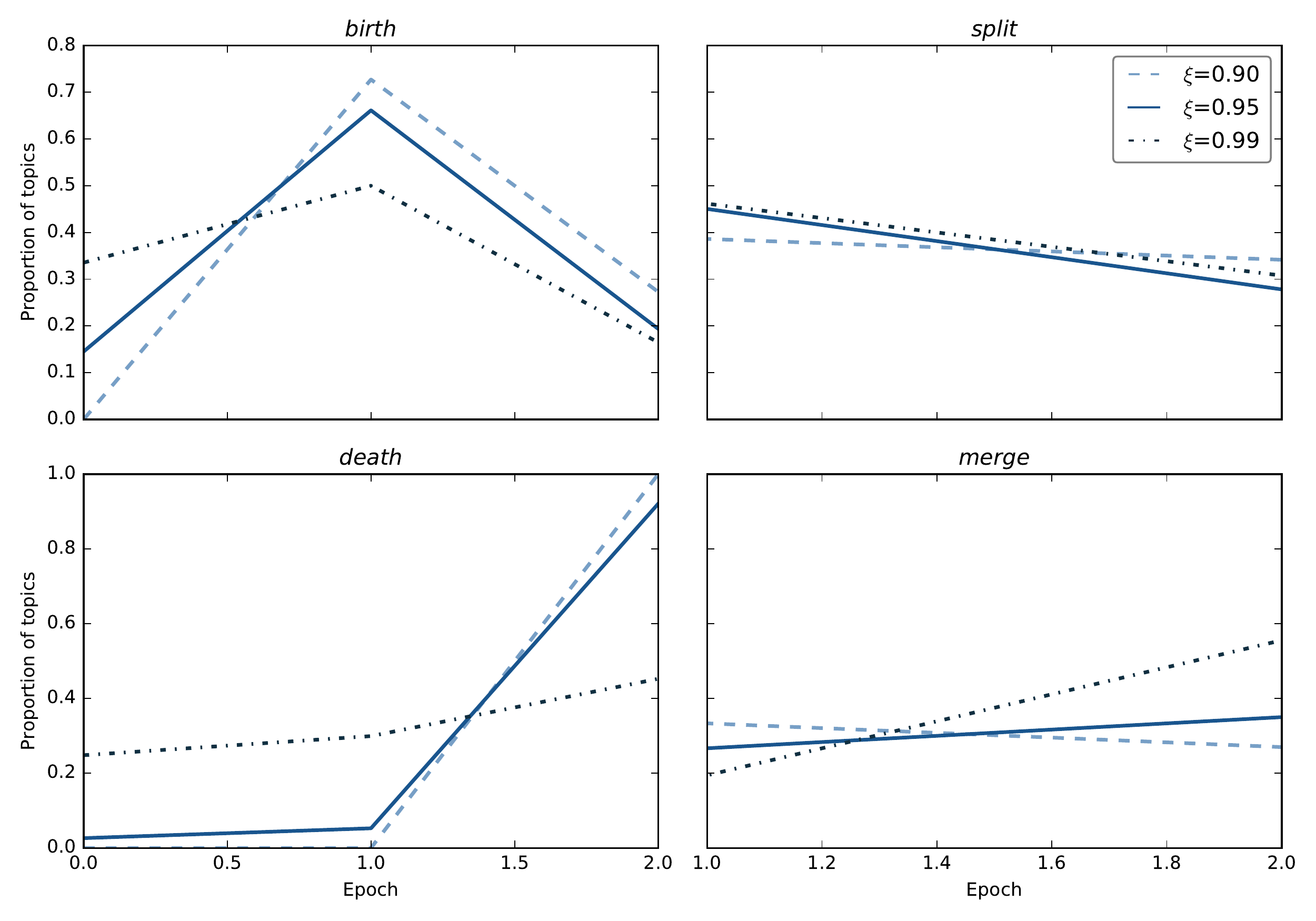}}
  \subfigure[Hellinger, 10 year epochs, 2 year overlap, ASD abstracts]{\includegraphics[clip, trim={0 0.4cm 0 0.1cm}, width=0.825\textwidth]{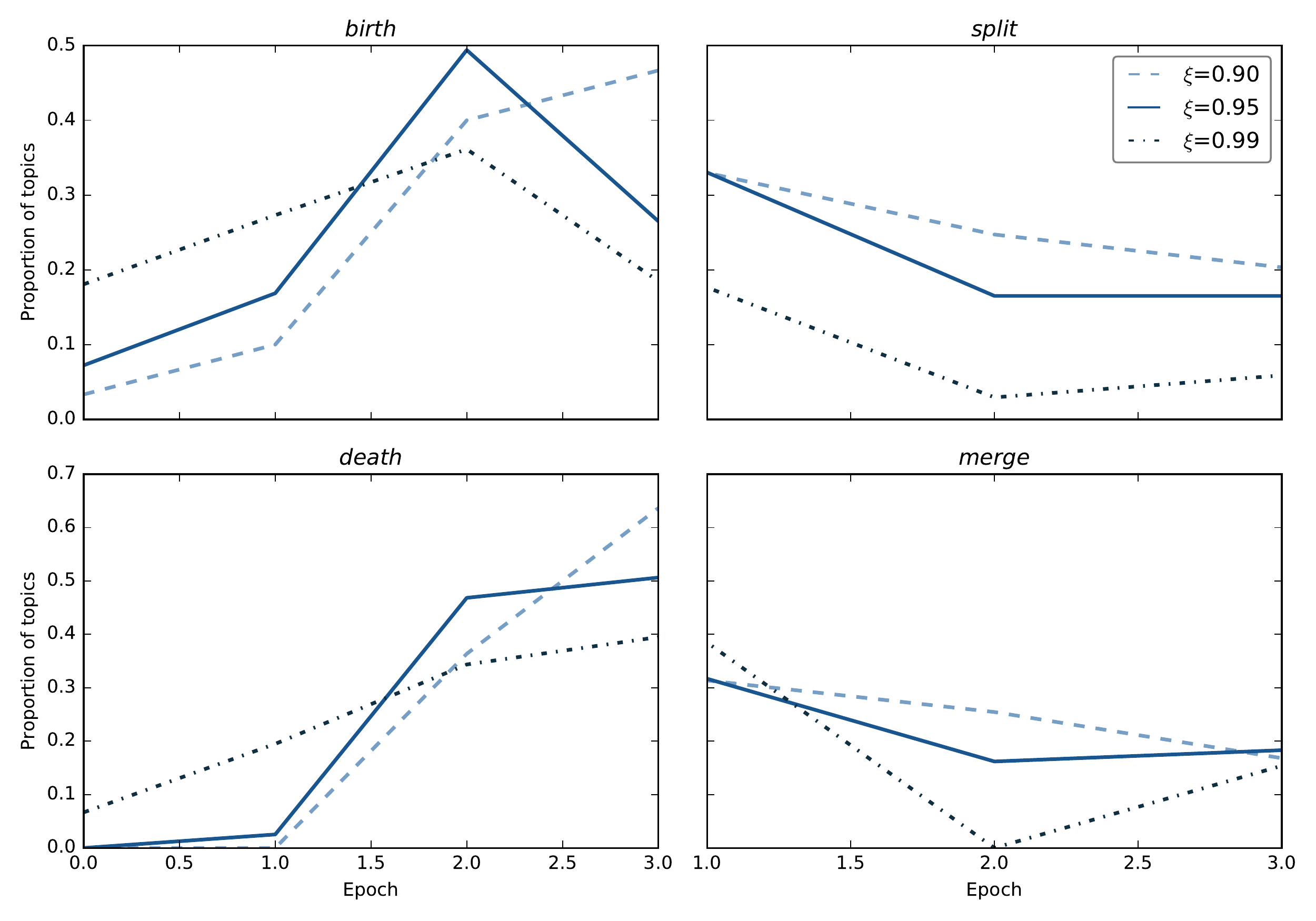}}
  \subfigure[Hellinger, 10 year epochs, 5 year overlap, ASD abstracts]{\includegraphics[clip, trim={0 0.4cm 0 0.1cm}, width=0.825\textwidth]{ASD_topic_life_per_epoch_L10O5_Hellinger_normalized.pdf}}
  \caption{The effect of the epoch overlap; ASD abstracts data set.}
  \label{f:comparisonOverlap}
\end{figure}

\paragraph{Successive epoch overlap}
Lastly we analysed the effect that the choice of successive epoch overlap has on the output of our algorithm. As in the preceding experiments we visualize a summary of the results using sets of plots of normalized rates of topic birth, death, splitting, and merging, each set corresponding to a different overlap -- please see Figure~\ref{f:comparisonOverlap}. Even a cursory examination of the plots readily reveals that the parameter in question has a profound effect. While a consistent pattern of changes can be observed in each set of plots, the most noticeable effect is on the rate of topic death. In particular it can be seen that as epoch overlap is reduced, the rate at which topics die off is increased. For example on our ASD data set using 10 year epochs, for the overlap of 5 years and the similarity CDF operating point $\zeta=0.95$ the steady state topic death rate is approximately 0.3. It increases to approximately 0.5 as the overlap is reduced to 2 years, and then to over 0.9 for no overlap at all. In other words, in the latter case over 90\% of topics exhibit no continuity of any sort (evolution, merging, or splitting), instead disappearing already within the epoch of their birth. Qualitatively this comes as no surprise -- the less overlap there is between successive epochs, the less relatedness can be expected between the sets of topics extracted in successive epochs. However, qualitatively, the magnitude of the effect is rather astonishing. Considering that most of the methods described in the literature which discretize time by epochs adopt the no-overlap design, our finding provides strong and valuable evidence that the performance of these methods could be improved with little effort, merely by a slight alteration in the manner discretization is performed.

\paragraph{Parameter combinations and topic life expectancy}
In the experiments so far we examined how the topic structure of a longitudinal document corpus and the evolution of this structure over time is affected by different free parameters of the proposed algorithm. Our results suggest that our algorithm is not very sensitive to the exact choice for the value of its parameters. This behaviour is highly desirable because it obviates the need for substantial amounts of data needed to learn sensible parameter values for a particular application. The one parameter which we found to be of particular importance is the amount of overlap between successive epochs used to discretize time. In particular we found that while the precise amount of overlap is not of particular importance, the introduction fo \emph{some} overlap (25--50\% of the epoch length) significantly improves the ability of our algorithm to capture temporal structural changes in the topic content of a corpus. This was most significantly demonstrated in the rate of topic death. As we noted earlier, without any epoch overlap most topics die off within the same epoch in which they are created. Hence we sought to investigate how the topic life expectancy is affected by different combinations of our algorithm's parameter values.

In the experiments described so far we examined the normalized topic birth and death rates independently. In other words we looked at the proportion of topics which are respectively newly created (born) and which disappear (die) as a proportion of the total number of topics extracted in the corresponding epoch. This information does not provide any insight into \emph{which} topics die off, that is, how long a specific topic has been in existence before it disappears. Here we adopt a more nuanced approach which comprises the following steps:
\begin{itemize}
  \item \textbf{Identification of topic creation:}\\ We identify the epoch in which a particular topic was created as the epoch of its birth \emph{or} the epoch in which the topic is created through the splitting or the merging of topics from the previous epoch.

  \item \textbf{Topic tracking:}\\ Following the creation of a topic we track it in the context of complex changes of its topic environment by considering its natural descendent to be its child in our temporal topic graph, with the highest degree of inter-topic similarity across all siblings. A topic is considered extant as long as it has any descendents.
  
  \item \textbf{Identification of topic death:}\\ Finally, we identify the epoch of a topic's death as the epoch in which the topic no longer has any descendants (children in our temporal topic graph).
\end{itemize}
Adopting this methodology we analyse the distribution of the life expectancy of a topic across our corpus and the effects of different algorithm parameters. Using the operating point corresponding to $\zeta=0.95$ on the inter-topic similarity CDF, we performed experiments using the six settings summarized in Table~\ref{t:paramsASD}. Our results are shown in Table~\ref{t:lifespanASD}.

\begin{figure}
  \centering
  \subfigure[]{\includegraphics[clip, trim={0 0.45cm 0 0}, width=0.75\textwidth]{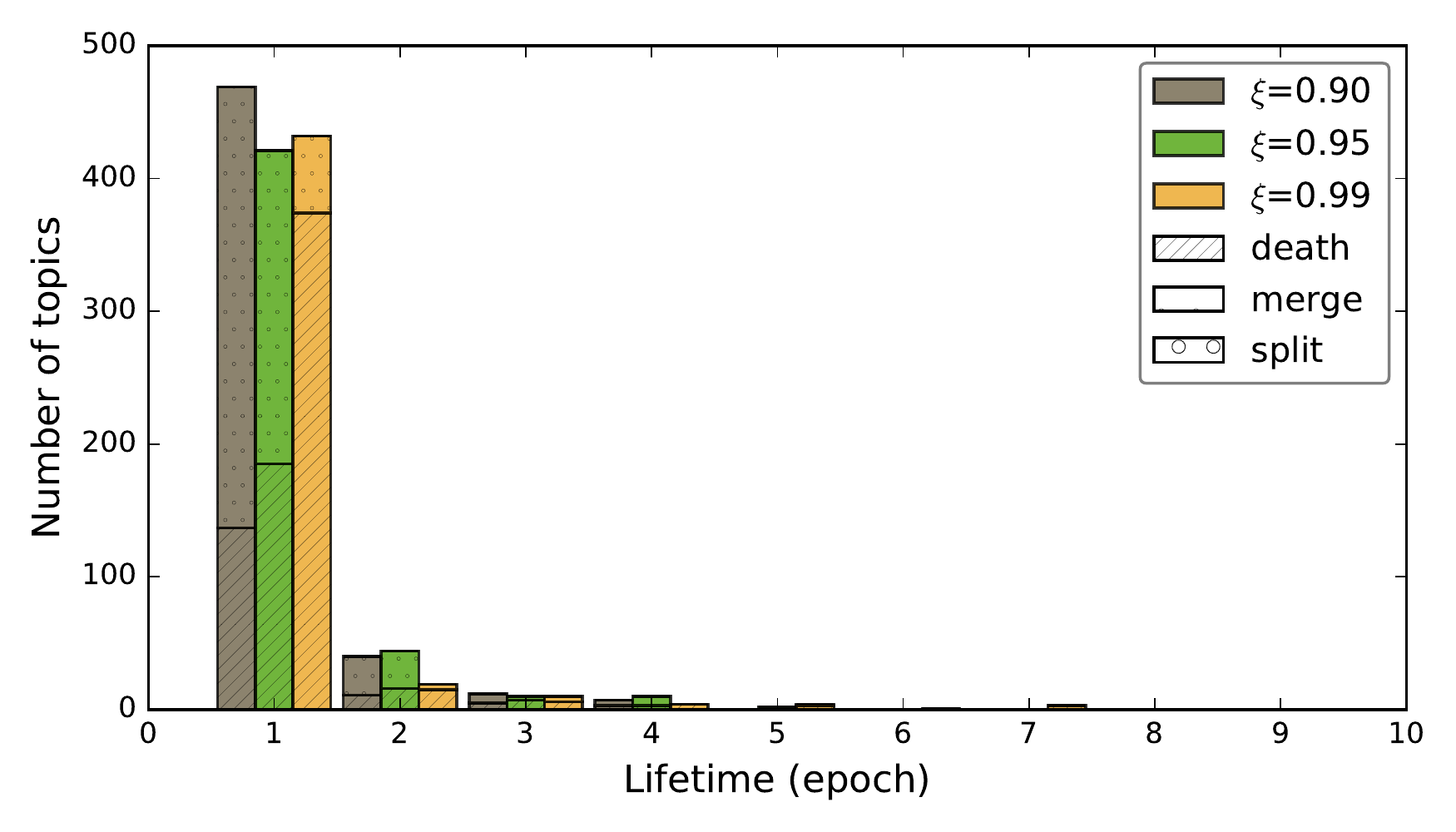}}
  \subfigure[]{\includegraphics[clip, trim={0 0.45cm 0 0}, width=0.75\textwidth]{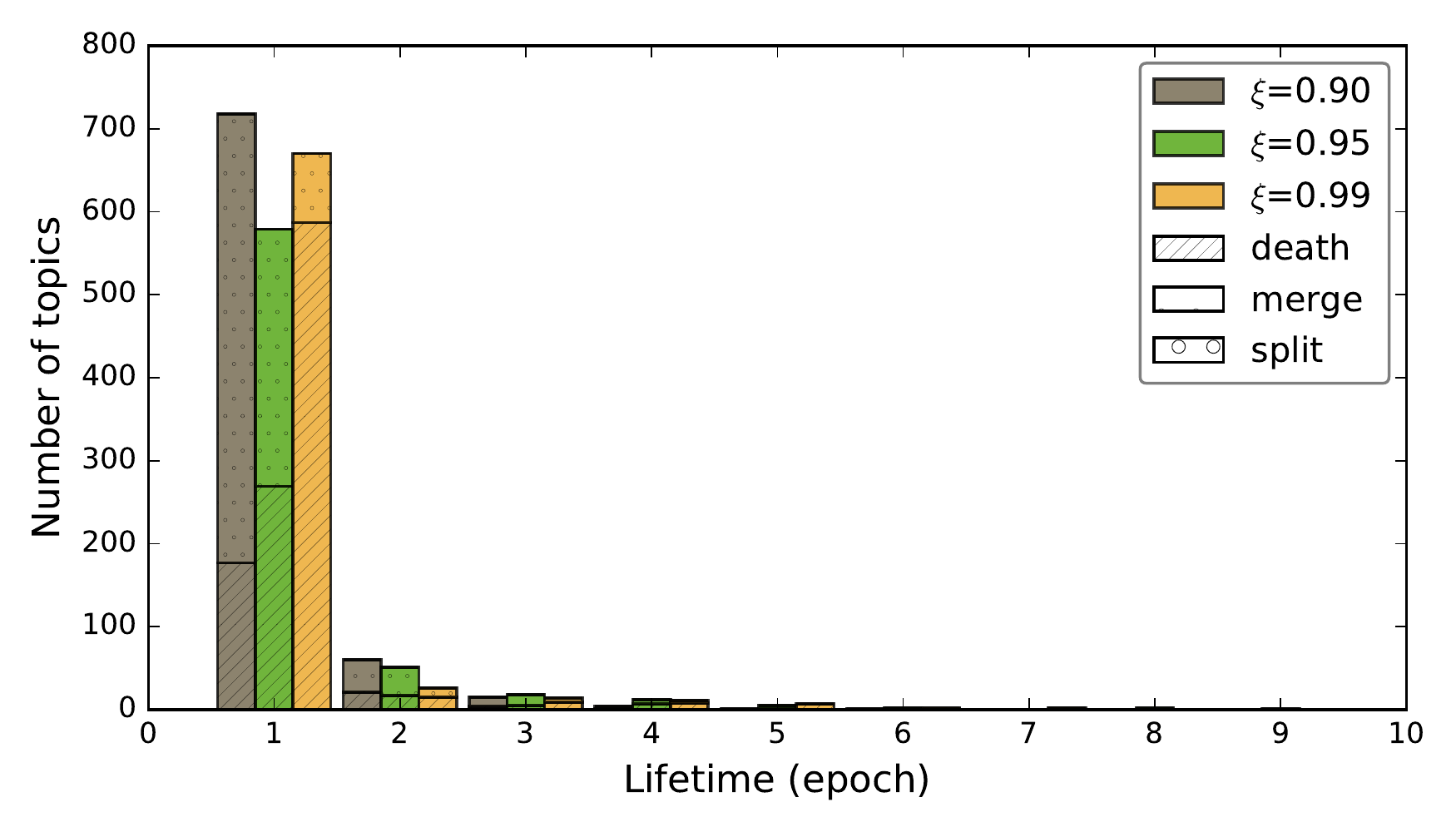}}
  \caption{A statistical summary of the manner in which topics cease to exist. The plots correspond to our ASD data set, analysed using the Bhattacharyya similarity measure, and (a) 10 year epochs with a 5 year overlap, and (b) 5 year epochs with a 2 year overlap.}
  \label{f:lifespanASD}
\end{figure}

Lastly, we examined the manner in which topics extracted by our algorithm cease to exist. In particular, for each new topic (where ``new'' in this context is taken to mean that a topic is either newly born, as defined previously, or that it is created by splitting or merging of topics from the previous epoch) we compute the time until it either dies (i.e.\ has no offspring in the following epoch), or splits or merges \emph{without any of its children} having it as the sole parent. The condition that none of a topic's children have the original topic as the sole parent is important because if there are such children, they can be seen to be evolved successors. If there are multiple children which have the ancestral topic as the sole parent then we track the lifetime of the topic down that path which leads to the longest topic lifetime. Our results are summarized in Table~\ref{t:lifespanASD} and the plots in Figure~\ref{f:lifespanASD}. In Table~\ref{t:lifespanASD} it is important to observe the lack of sensitivity of our method to its specific settings. From the plots in Figure~\ref{f:lifespanASD} it is interesting to notice that most of the topics cease to exist already in the first epoch. This observation provides potentially useful insight into the choice of the temporal analysis scale. As expected, increasing the pruning threshold $\zeta$, which has the effect of decreasing inter-epoch topic linkage, results in an increased proportion of outright topic deaths. Lastly, it is insightful to observe that following the first epoch, for a fixed specific pruning threshold the proportion of topic deaths decreases.

\begin{table}
  \centering
  \setlength{\tabcolsep}{0.5cm}
  \caption{Average lifespan of topics (in epochs) extracted from our autism spectrum disorder abstracts data set. $^*$For full detail of different settings see Table~\ref{t:paramsASD}. }
  \vspace{8pt}
  \begin{tabular}{lccc||ccc}
    \Hline
    Epoch length       & \multicolumn{3}{c}{10 years} & \multicolumn{3}{c}{5 years}\\
    Setting number$^*$            & 1         & 2        &  3      & 4         & 5        &  6\\
    \fline  
    Hellinger       &  1.28   &  1.27  & 1.21  &  1.38  &  1.30  & 1.27\\
    Bhattacharyya   &   1.31  &  1.30  & 1.22  &  1.39  &  1.31  & 1.26\\ 
    Pseudo-Jaccard  &   1.30  &  1.23  & 1.15  &  1.23  &  1.28  & 1.17\\
    \Hline
  \end{tabular}
  \label{t:lifespanASD}
  \vspace{0pt}
\end{table}

\subsubsection{Qualitative results}\label{sss:qual}
In the previous section we described an extensive set of experiments providing insight into the role that different free parameters of our algorithm have on its output. Importantly we demonstrated that within a wide range of what may be described as reasonable choices for the parameter values, the proposed method exhibits a high degree of robustness. Following these encouraging quantitative findings, using our (human) higher level semantic understanding of the corpora used, we now examine if the output of our algorithm is meaningful and ultimately useful.

\paragraph{Case study 1: ASD and genetics.\label{sub:Case-study-1}}
While the exact aetiology of the ASD is still poorly understood, the existence of a significant genetic component is beyond doubt~\cite{Mile2011}. Work on understanding complex genetic factors affecting the development of autism, which possibly involve multiple genes which interact with each other and the environment, is a major theme of research and as such a good case study on which the usefulness of the proposed method can be illustrated.

\begin{figure}
  \centering
  \includegraphics[width=0.95\textwidth]{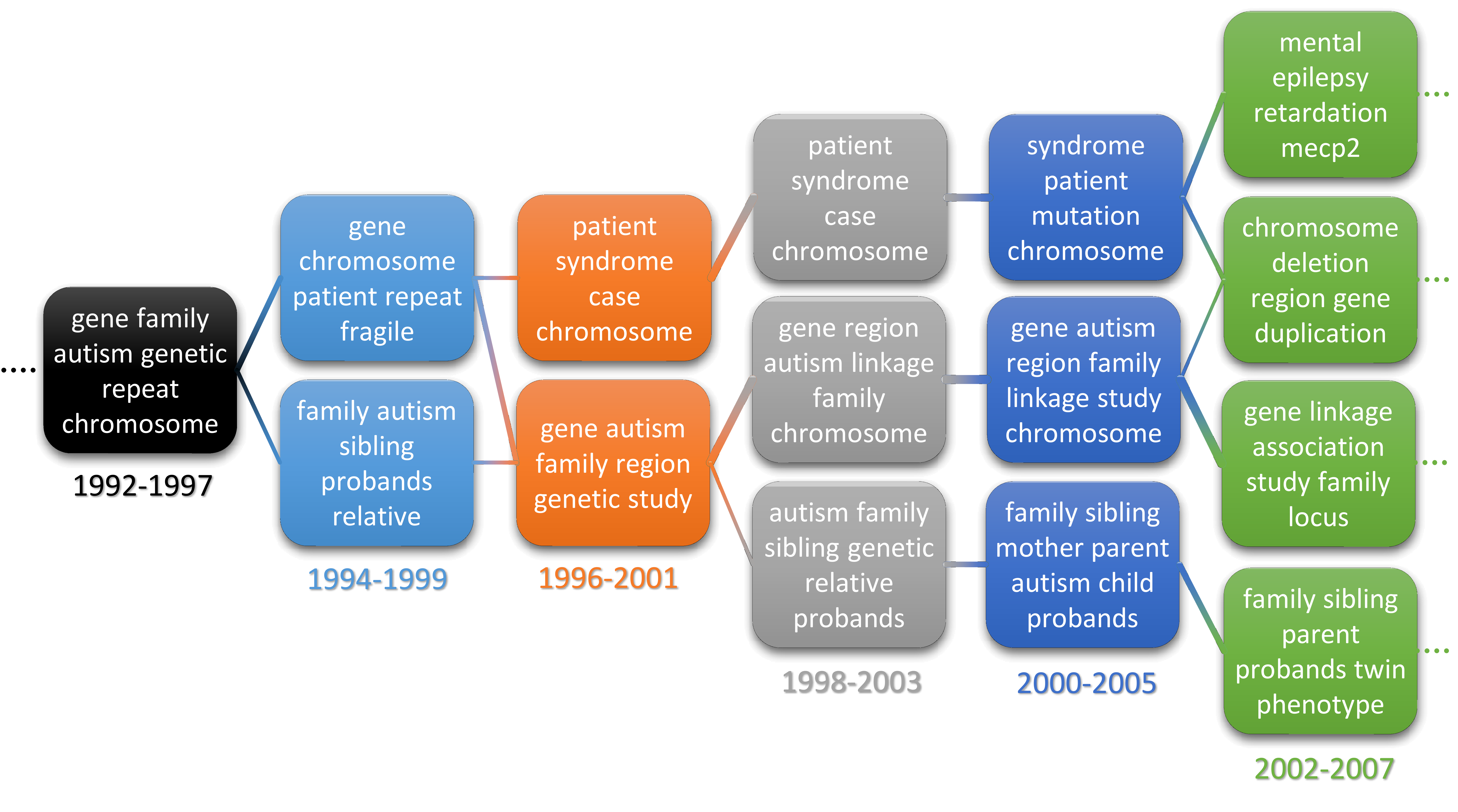}
  \caption{Dynamics of the topic most closely associated with the concept of ``genetics''. A few dominant words are shown for each topic (shaded boxes).}
  \label{fig:Dynamics-of-ASD}
\end{figure}

We started by identifying the topic of interest as that with the highest probability of the terms ``gene'' or ``genetic'' conditioned on the topic, and tracing it back in time to the epoch in which it originated. This led to the discovery of the relevant topic in the epoch spanning the period 1986--1991. Figure~\ref{fig:Dynamics-of-ASD} shows the evolution of this topic from 1992 revealed by our method (due to space constraints only the most significant parts of the similarity graph are shown; minor changes to the topic before 1992 are also omitted for clarity, as indicated by the dotted line in the figure). Each topic is labelled with its first few dominant terms. The following interpretation of our findings is readily apparent. Firstly, in the period 1992--1997, the topic is rather general in nature. Over time it evolves and splits into topics which concern more specific concepts (recall that such splitting of topics cannot be captured by any of the existing methods). For example by the epoch 2002--2007 the single original topic has evolved and split into four topics which concern the following:
\begin{itemize}
  \item The relationship between mutations in the gene \texttt{mecp2} (essential for normal functioning of neurone), and mental disorders and epilepsy  (it is estimated that one third of ASD individuals also have epilepsy),\\[-5pt]
  \item Gene alternations, for example the duplication of \texttt{15q11--13} and deletion of \texttt{16p11.2} both of which are associated with ASD,\\[-5pt]
  \item Genetic linkage association analysis and heritability of autism, and\\[-5pt]
  \item Observational work on autistic twins and probands with siblings on the spectrum.
\end{itemize}

Our framework also allows us to look `back' in time. For example, by examining the topics that the 1992 genetics topic originate from we discovered that the topic evolved from the early concept of ``infantile ASD''~\cite{kanner1946irrelevant}.

\paragraph{Case study 2: ASD and vaccination.\label{sub:Case-study-2}}
For our second case study we chose to examine research on the relationship between ASD development and vaccination. This subject has attracted much attention both in the research community, as well as in the media and the general public. The controversy was created with the publication of the work by Wakefield~\cite{WakeMurcAnth1998} which reported epidemiological findings linking MMR vaccination and the development of autism and colitis. Despite the full retraction of the article following the discovery that it was fraudulent, and numerous subsequent studies who failed to show the claimed link, a significant portion of the general public remains concerned with the issue.

\begin{figure}
  \centering
  \includegraphics[width=1\textwidth]{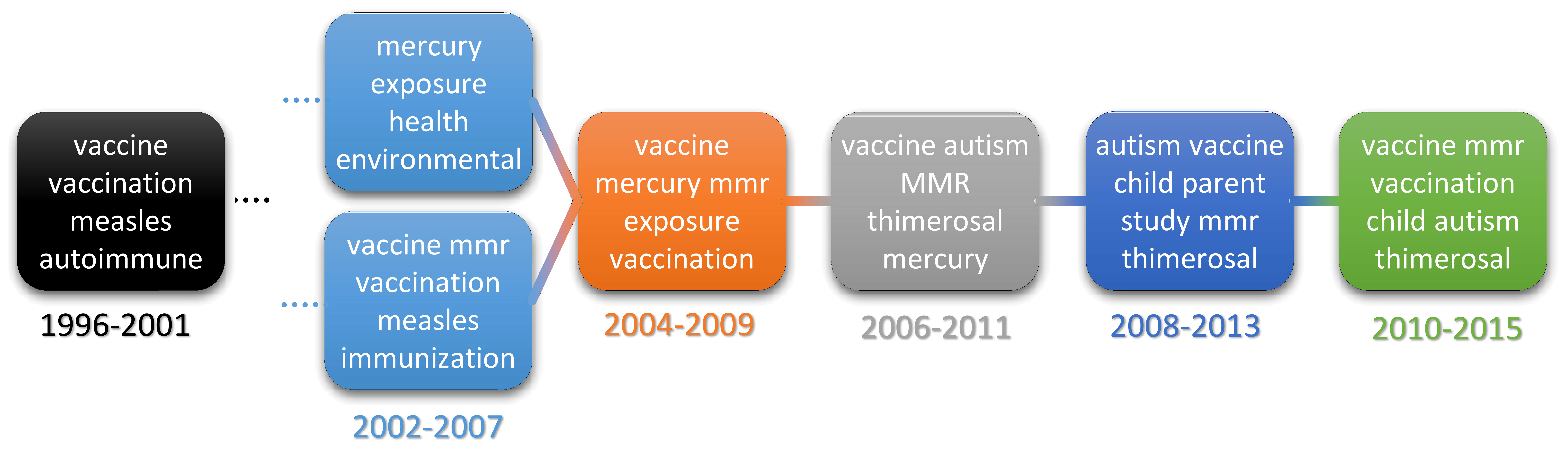}
  \caption{Dynamics of the topic most closely associated with the concept of ``vaccination''. Notwithstanding the rejection of any link between vaccination and autism, this topic remains active albeit in a form which evolved over time.}
  \label{fig:Dynamics-of-vaccination}
\end{figure}

As in the previous example, we begin by identifying the topic with the highest probability of the terms ``vaccine'' and ``vaccination'' conditioned on the topic, and tracing it back to the epoch in which it first emerged. Again, a single topic was readily identified, in the epoch spanning the period 1996--2001. Notice that this is consistent with the publication date of the first relevant publication by Wakefield~\cite{WakeMurcAnth1998}. The evolution of the topic is illustrated in Figure~\ref{fig:Dynamics-of-vaccination} in the same way as in the previous section. It can be seen that the original topic concerned the subjects initially brought to attention such as ``measles'', ``vaccine'', and ``autism''. In the subsequent epoch, when the original claim was still thought to have credibility, the topic evolves and splits into numerous others mirroring research directions taken by various researchers. Following this period and the revelations of its fraudulence, the topic assumes mainly single-threaded evolution, at times incorporating various originally separate ideas. For example observe the independent emergence of the term ``mercury''. Though initially unrelated to it this topic merges with the topic that concerns vaccination which can be explained by the widely publicized thiomersal (vaccine preservative) controversy (again note that such merging of topics cannot be captured by the existing methods). Although rejected by the medical community due to a lack of evidence, this topic can be seen as persisting to date.

\paragraph{Case study 3: MetS and plasma fatty acids.}
As noted earlier MetS is highly associated with the risk of developing type 2 diabetes and cardiovascular diseases, and is characterized by insulin resistance, abdominal obesity, and high blood pressure, all of which are intimately linked with dyslipidemia and elevated plasma fatty acid levels. In this case study we sought to investigate patterns associated with topics concerning this aspect of MetS. 

\begin{figure}
  \centering
  \includegraphics[width=1\textwidth]{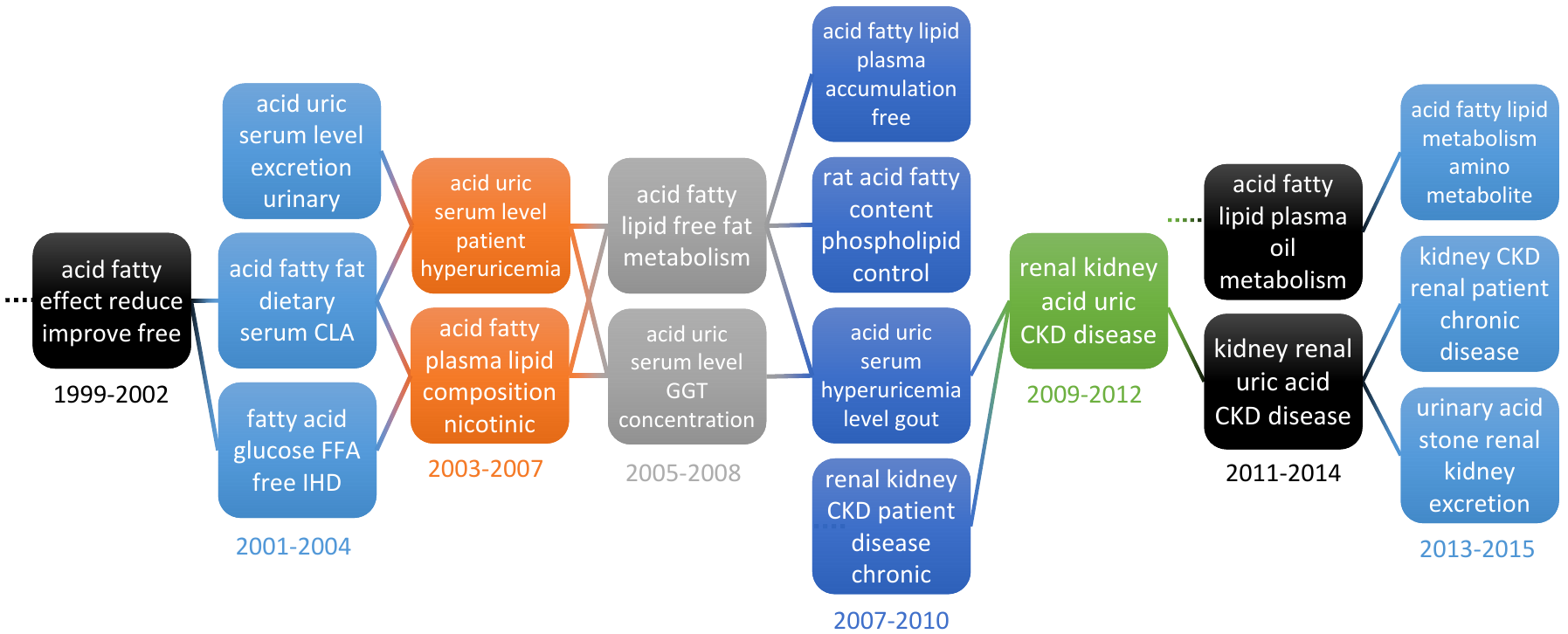}
  \caption{Dynamics of the topic most closely associated with the concept of fatty acids (i.e.\ in the context of our vocabulary the terms ``acid'' and ``fatty''). Fatty acid metabolism plays a key role in the metabolic syndrome.}
  \label{f:caseMetS}
\end{figure}

As in the previous two case studies we began by identifying the topics with the highest probability of the relevant terms (in this case ``acid'' and  ``fatty'') conditioned on the topic, and tracing it back to the epoch in which it first emerged. For the sake of clarity of visualization we focus on the period starting with the 1999--2002 epoch which shows some of the most interesting dynamics. The evolution of topic of interest in this period is shown in Figure~\ref{f:caseMetS}. Examples of illustrative observations based on this section of our similarity graph include the following:
\begin{itemize}
  \item The 2001--2004 epoch gives birth to a topic dominated by the terms ``acid'' and ``uric''. This topic merges a topic concerned with dietary fat, thereby resulting in a topic associated with hyperuricemia in the following epoch. Both the semantics of the extracted topics themselves as well as the aforementioned dynamics are readily interpreted from and supported by the literature on MetS -- the relationship between dietary fat (and types thereof) intake and and plasma uric acid levels has been a topic of significant research interest~\cite{HudgHellSeidNees+1996}.\\[-5pt]
  
  \item The broad topic concerning fatty acid metabolism present in the 2005--2008 epoch can be seen to branch into three distinct research directions on the accumulation of free fatty acids, controlled clinical trials in murine models, and hyperuricemia and gout. This pattern is again congruent with the actual development of research in the field in recent years: the role of free fatty acid accumulation has been attracting an increasing amount of research attention~\cite{SeppVehkHakkGoto+2002}, murine models have been widely used to study specific aspects of MetS in controlled conditions unfeasible with human subjects~\cite{MackQuarVerrMack+2006}, and there is an accumulating body of evidence supporting a causal link between MetS and the increasing incidence of gout (historically known as ``the rich man's disease'') in the Western world~\cite{ChoiFordLiCurh2007}. 
\end{itemize}

\section{Summary and conclusions\label{s:conc}}
%In this paper we described an extension of the method introduced in our previous work, for tracking the topic structure of a longitudinal document corpus across time and in the presence of complex changes and topic interactions.
This paper focused on the problem of modelling and extracting the topic structure of a longitudinal document corpus over time. The approach we described starts with a discretization of time into epochs which may overlap. Then, using the approximation that the topic structure within each epoch is temporally locally static, the aforesaid structure is modelled and extracted using a hierarchical Dirichlet process. Finally, the evolution of the topic structure over time is captured using a temporal graph underlain by an inter-topic similarity measure. The graph, initially populated by edges between all pairs of topics in two consecutive epochs is pruned automatically and the result used to infer complexity structural changes over time which the existing methods in the literature cannot model, including the emergence and disappearance of topics and their evolution over time, as well as the merging and splitting of an arbitrary number of topics.

The proposed framework was evaluated extensively on two large real-world data sets of abstracts of scientific papers, one concerning the autism spectrum disorder and the other the metabolic syndrome. This data was collected by ourselves, and made free for public use. Our detailed quantitative analysis of the effects that the free parameters of the proposed method have on its performance revealed a number of important insights. We found that within a wide range of parameter values our algorithm was little affected by the specific value choices. Another important finding, the significance of which extends further than the scope of the proposed algorithm, is that in the discretization of time into epochs it is important that successive epochs overlap. The significantly inferior performance observed with non-overlapping epochs has immediate consequences for the interpretation of previous work and the findings reported in the literature, suggesting a simple and immediate way of enhancing the performance of any algorithm which did not adopt the use of overlapping epochs. Lastly, on several case studies highly relevant to the currently popular directions of research on ASD and MetS, our algorithm's output was analysed qualitatively, and shown to capture well the actual developments in these fields.

\bibliographystyle{plain}
\bibliography{../KAIS,../../../my_bibliography,../../../oa_physiology}

\correspond {Adham Beykikhoshk, Pattern Recognition and Data Analytics Centre, Deakin University, Australia. E-mail: abeyki@deakin.edu.au}
\end{document}